\documentstyle[12pt,aaspp4,amstex]{article} 

\def\pht{\tiny {\mbox{H,tot}}}
\def\hh{\tiny {\mbox{HI}}}
\def\hhh{\tiny {\mbox{HII}}}
\def\phet{\tiny {\mbox{He,tot}}}

\def\heh{\tiny {\mbox{HeII}}}
\def\hehh{\tiny {\mbox{HeIII}}}	

\def\ltsima{$\; \buildrel < \over \sim \;$}
\def\simlt{\lower.5ex\hbox{\ltsima}}
\def\gtsima{$\; \buildrel > \over \sim \;$}
\def\simgt{\lower.5ex\hbox{\gtsima}}

\begin{document}

\title{Cosmic Reionisation by Stellar Sources: Population II Stars}

\author{Aaron Sokasian\altaffilmark{1}, Tom Abel\altaffilmark{2},
Lars Hernquist\altaffilmark{1}, and Volker Springel\altaffilmark{3}}

\altaffiltext{1}{Harvard-Smithsonian Center for Astrophysics, 60
Garden Street, Cambridge, MA 02138, USA} \altaffiltext{2}{Department
of Astronomy and Astrophysics Penn State University, 525 Davey Lab,
University Park, PA 16802, USA} \altaffiltext{3}{Max-Plank-Institut
f$\ddot{\text{u}}$r Astrophysik, Karl-Schwarzschild-Stra$\beta$e, 1,
85740 Garching bei M$\ddot{\text{u}}$nchen, Germany}

\authoremail{asokasia@cfa.harvard.edu}

\begin{abstract}

We study the reionisation of the Universe by stellar sources using a
numerical approach that combines fast 3D radiative transfer
calculations with high resolution hydrodynamical simulations. By
supplementing a {\it one-step} radiative transfer code specifically
designed for following ionisation processes with an adaptive
ray-tracing algorithm, we are able to significantly speed up the
calculations to the point where handling a vast number of sources
becomes technically feasible.  This allows us to study how dim
low-mass sources, excluded in previous investigations owing to
computational limitations, affect the morphological evolution of the
reionisation process.

Ionising fluxes for the sources are derived from intrinsic star
formation rates computed in the underlying hydrodynamical simulations.
Analysis of numerically converged results for star formation rates and
halo mass functions allows us to assess the consequences of not
including low-mass objects and enables us to correct for resolution
effects.  With these corrections, we are able to reduce the effective
mass resolution limit for sources to $M\sim4.0 \times 10^{7}\ h^{-1}$
M$_{\odot}$, which is roughly an order of magnitude smaller than in
previous studies of this kind.  Our calculations reveal that the
process by which ionised regions in the IGM percolate is complex and
is especially sensitive to the inclusion of dim sources.  Moreover, we
find that given the same level of cosmic star formation, the number of
ionising photons required to reionise the Universe is significantly
overestimated if sources with masses below $\sim10^9 \ h^{-1}$
M$_{\odot}$ are excluded.  This result stems from the fact that
low-mass sources preferentially reside in less clumpy environments
than their massive counterparts.  Consequently, their exclusion has
the net effect of concentrating more of the cosmic ionising radiation
in regions which have higher recombination rates.

We present the results of our reionisation simulation assuming a range
of escape fractions for ionising photons and make statistical
comparisons with observational constraints on the neutral fraction of
hydrogen at $z\sim6$ derived from the $z=6.28$ SDSS quasar of Becker
and coworkers.  We find that given the amplitude and form of the
underlying star formation predictions, an escape fraction near
$f_{esc}=0.10 - 0.20$ is most consistent with the observational
results. In these models, reionisation is expected to have occurred
between $z\sim7 -8$, although the IGM remains fairly opaque until
$z\simeq6$. 

Our method is also capable of handling the simultaneous reionisation
of the helium component in the IGM, allowing us to explore the
plausibility of the scenario where sources with harder spectra are
primarily responsible for reionisation. In this case we find that if
the sources responsible for reionising hydrogen by $z\sim8$ had
spectra similar to AGNs, then the helium component of the IGM should
have been reionised by $z\sim6$. We find that such an early
reionisation epoch for helium does not necessarily conflict with
observational constraints obtained at $z\simeq3$, but may be
challenged by future observations at higher redshifts.

The recent WMAP measurements of the electron scattering optical depth
($\tau_e=0.17 \pm 0.04$ according to the ``model independent''
analysis of Kogut et al.) appear to be inconsistent with the
relatively late onset of reionisation by the normal population II type
stars that we consider. In order to simultaneously match the
observations from the $z=6.28$ SDSS quasar and the optical depth
measurement from WMAP with the sources modeled here, we require an
evolving escape fraction that rises from $f_{esc}=0.20$ near
$z\simeq6$ to $f_{esc}\gtrsim10$ at $z\sim18$.  Such a steep
enhancement in the stellar production rate of ionising photons would
be consistent with an IMF that becomes more and more top heavy
with increasing redshift.

\end{abstract}

\keywords{radiative transfer -- diffuse radiation -- intergalactic
medium -- galaxies: quasars}

\section{INTRODUCTION}

The nature of the sources responsible for ionising the neutral
hydrogen in the intergalactic medium (IGM) has been the subject of a
long-standing debate.  There are both observational and theoretical
indications that some or perhaps all of this ionising background comes
from stars and not hard sources, such as quasars. While it seems clear
that quasars provide a significant contribution to the ionising
background at low redshifts, $z\sim 3$, the steep drop in the
abundance of bright quasars at earlier times brings into question
whether they are capable of maintaining the ionised state of the IGM
at $z\simgt 4-5$.  In an earlier paper (Sokasian et al. 2003), we
showed that the observed opacities in both H {\small I} and He {\small
II} indicate that quasars alone cannot produce the required
emissivity to match observations at $z\simeq4$.  The situation is
exacerbated at higher redshifts where observations by Fan et
al. (2000), Becker et al. (2001), and Djorgovski et al. (2001) imply
that the IGM was highly ionised at $z\simeq6$ even though the
emissivity from bright quasars was likely unimportant.  Given this
discrepancy, it appears that a substantial portion of the ultraviolet
photons at $z>4$ were produced by an additional class of sources.

It has been argued that star forming galaxies may yield the required
ionising emissivities (e.g. Haehnelt et al. 2001, Bianchi et al. 2001,
Steidel et al. 2001), although it is possible that harder
low-luminosity active galactic nuclei (AGN) may also make up the
difference (Haiman \& Loeb 1998).  Early star formation also provides
a possible explanation for the widespread abundance of metals in the
IGM (Cowie et al. 1995) while reionisation by faint AGNs may provide a
more plausible solution if the escape fraction for stellar radiation
from galaxies is restrictively small (Schirber \& Bullock 2002, Wood
\& Loeb 2000).

There have been numerous studies of cosmological reionisation using
semi-analytical methods (e.g. Haardt \& Madau 1996, Madau et al. 1999,
Chiu \& Ostriker 2000, Valageas \& Silk 1999). In these models, the
impact of the UV background is treated using a spatially and
directionally averaged radiative transfer equation which incorporates
the mean luminosity function per unit volume for the source term and
the statistical properties of absorbing clouds for the recombination
term. Inevitably though, to accurately address questions related to
the time-dependent propagation of radiation fronts, especially during
the early stages of the reionisation process where the radiation field
is highly inhomogeneous, a detailed radiative transfer approach is
required.  This has proved to be an extremely challenging endeavour
owing to the high dimensionality of the problem.  A brief review of
the problem and the various techniques which have been employed in
recent years is presented in Razoumov et al. (2002).

One of the major challenges involved with such exercises is
determining how to efficiently process the large number of sources
that may be responsible for reionisation. In the case of bright
quasars, the associated number density is small enough that
straightforward approaches can be employed (e.g. Sokasian et al. 2001,
2002). The situation becomes much more difficult when galaxies are the
primary sources. For typical cosmologies, a $10^3$ Mpc$^3$ comoving
volume can easily harbour $>10^4$ star forming galaxies by $z=6$.

Tracking the ionising flux from such a large pool of sources over many
time steps inevitably requires the implementation of sophisticated
algorithms designed to efficiently process the cumulative effects from
each source. By combining a photon conserving algorithm with two
independent hierarchies of trees, one for rays (Abel \& Wandelt 2002)
and one for the sources themselves, Razoumov et al. (2002) were able
to improve the efficiency of the calculations to study the
inhomogeneous process of reionisation from star forming sources down
to a redshift of $z=4$.  In particular, they were able to carefully
follow time-dependent radiative transfer calculations on a uniform
$128^3$ Cartesian grid representing a volume with a comoving length of
$L\sim 7h^{-1}$ Mpc on each side which stored the results of
pre-computed hydrodynamical simulations of galaxy formation. As a
result of their calculations, the authors were able to address a
number of questions related to the morphological evolution of the
reionisation process. In particular, they showed a picture of stellar
reionisation in which photons initially do not travel far from
ionising sources, in contrast to images from simulations by Gnedin
(2000).

More recently, Ciardi et al. (2003b) combined high-resolution N-body
simulations and a semi-analytic model of galaxy formation with a Monte
Carlo radiative transfer code to study reionisation in a volume with a
comoving length of $20 h^{-1}$ Mpc. In their analysis they assessed
the effect of the environment by performing a subsequent simulation on
a smaller $10 h^{-1}$ Mpc comoving box centered on a cluster.  They
found that the environment where the radiation is produced modifies
reionisation through the influence of the density dependence of the
recombination rates. This conclusion raises the question of whether
the exclusion of the copious number of dim low-mass sources which fall
below the resolution limit of these simulations may play a more
important role in determining the morphological evolution of the
reionisation process. The fact that most models of galaxy formation
predict that the bulk of the resultant ionising radiation is produced
in relatively massive systems has provided some motivation for
ignoring the low-mass sources altogether, especially in lieu of the
high computational cost of radiative transfer calculations.  However,
since dim low-mass sources reside in relatively less clumpy
environments compared to their brighter counterparts, their overall
contribution to the ionising emissivity in the IGM may be significant
in spite of their intrinsic faintness.

We examine the impact of low-mass sources by conducting very fast
radiative transfer calculations capable of processing a large number
of sources from high resolution hydrodynamic simulations.  In
particular, by supplementing our previous {\it one-step} radiative
transfer code (see Sokasian et al. 2001) with an adaptive ray tracing
algorithm we are able to significantly speed up the calculations to
the point where handling more than $10^5$ sources becomes technically
feasible.  Our cosmological simulation is taken from a high resolution
run in a series of calculations performed by Springel \& Hernquist
(2003a), and utilises $2\times 324^3$ particles in a box of comoving
length $L\sim 10 h^{-1}$ Mpc.  Radiative transfer calculations are
then performed on a 200$^3$ grid where we store the corresponding density
fields at each redshift. This allows us to simulate reionisation with
a source mass resolution that is about an order of magnitude below the
limit of the Ciardi et al. (2003b) simulations.  Furthermore, the use
of a 200$^3$ grid offers a factor of $\sim4$ improvement in
spatial resolution for the same box length. It is also important to
point out that by using hydrodynamics to describe the gas physics,
rather than an N-body simulation, we are able to estimate the clumping
on unresolved scales in our grid by computing the spatial distribution
of particles within a given cell (see Sokasian et al. 2001). This
represents an important advantage over the density interpolation
schemes associated with N-body codes which smooth out density
variations on sub-grid scales.  As a result we expect to more properly
account for the recombination rates in high density regions where
clumping is important.

In the present paper, which is part of a series examining
reionisation, we focus on the redshift interval $z\simeq20$ to
$z\simeq6$ with specific interest in how the level of cosmic star
formation predicted in the comprehensive set of hydrodynamic
simulations conducted by Springel \& Hernquist (2003a) performs in the
context of observational constraints on the opacity of the IGM at
$z\simeq6$.  Our method also treats the simultaneous reionisation of
helium in the IGM, allowing us to explore the plausibility of the
scenario where AGNs are primarily responsible for reionisation.

Here, we restrict our attention to sources of radiation similar to
normal, star-forming galaxies at the present day.  In particular, we
do not account for massive, population III stars that may have been
prevalent at $z\simgt 20$.  Tentative measurements of polarisation in
the CMB by the WMAP satellite (Kogut et al. 2003) suggest that much of
the IGM was ionised earlier than indicated by the SDSS quasars (e.g.,
Spergel et al. 2003).  We explore the implications of this result in
the context of our present model by allowing the escape fraction to
evolve with redshift.  In the future, we will explore the combined
effect of reionisation by both population II and population III stars.

In the next section we review some of the technical aspects related to
our method, including the approximations inherent to our radiative
transfer calculations. We then follow with a description of the
cosmological simulation that will be used in our analysis.  Next, we
discuss our methodology for source selection and present an in-depth
analysis regarding resolution effects and how to correct for them.
Finally, we discuss the results of our reionisation simulation and
review the properties of the inhomogeneously ionised IGM and make
comparisons with observational constraints.

\section{SIMULATING COSMOLOGICAL REIONISATION}

In general, a description of the evolution of ionised regions around
cosmological sources requires a full solution to the radiative
transfer equation.  Such a solution would yield everywhere the
monochromatic specific intensity of the radiation field in an
expanding universe: $I_{\nu}\equiv I(t,\vec{x},\hat{n},\nu)$, where
$\hat{n}$ is a unit vector along the direction of the propagation of a
ray with frequency $\nu$.  Presently, it is computationally
impractical to obtain a complete, multi-dimensional solution for
$I_{\nu}$ at the high resolution required for cosmological
simulations.  Our approach relies on an algorithm which approximates
the full solution from ionising sources by semi-independently
processing the net effect of each source separately (see Sokasian et
al. 2001).  Owing to the large number of sources and the fact that the
order in which sources are processed is randomised at each step, the
departure from the fully self-consistent solution is negligible.  (We
have verified this explicitly by running different cases in which the
ordering of the sources varies and shown that our results are
unaffected.)  By adopting this approximation, we are able to employ a
simple jump condition to compute all radiative ionisations from a
given source in a single step. This eliminates the need to repeatedly
re-cast rays and calculate rates at every time step, thereby greatly
speeding up the analysis.  Ionisation fractions within ionised regions
are then computed by assuming photoionisation equilibrium.

To further speed up the calculations, the ray casting scheme used in
Sokasian et al. (2001) has been modified to incorporate an adaptive
scheme based on the method described by Abel \& Wandelt (2002). In
particular, the transfer of direct source photons is computed on a
uniform Cartesian grid by casting a set of base rays which
subsequently split hierarchically according to a local refinement
criterion which involves keeping the ratio of the area of a cell to
the area associated with the ray on a given hierarchical level above a
fixed threshold value. This criterion ensures that a minimum number of
rays pass through any given cell. The underlying pixelisation scheme
that governs how rays are split has the property of exactly tiling a
sphere with equal areas, ensuring equal photon flux per ray for
isotropic sources. Rays are propagated until they travel at most
$\sqrt{3}$ times the box length (assuming periodic boundary
conditions) or the assigned number of photons is exhausted.  Once the
ionisation level in the simulation box has reached significant levels,
periodic boundary conditions are turned off and photons reaching the
edges are added to a pool associated with the diffuse background.  At
the end of each interval, the effect of a diffuse ionising background
is accounted for by casting rays of ionising radiation (according to a
detailed prescription) inward from the sides of the simulation box.
Density fields, clumping factors and information regarding sources is
specified from outputs at desired redshifts from the underlying
cosmological simulation (see Sokasian et al. 2001 for technical
details).

Our approach entails a number of straightforward approximations to
simplify the calculations; we summarise them briefly here.  First, our
radiative transfer calculations are done on a uniform grid whose scale
$L$ will always be much smaller than the horizon, $c/H(t)$, where $c$
is the speed of light and $H(t)$ is the time dependent Hubble
constant.  This eliminates the need to include Doppler shifting of
frequencies in line transfer calculations.  Additionally, if the light
crossing time, $L/c$, is much shorter than the ionisation timescale,
the time dependence of the intensities drops out as well.  In the
volumes we simulate, this will certainly be true and so we make this
approximation.  Next, we assume that the density field will experience
negligible cosmological evolution during a single timestep.  This
assumption allows us to perform all our radiative calculations during
a given timestep in a static density field, greatly reducing the
complexity of the algorithm.  In our analysis we find that a constant
time step of $4.0\times10^7$ yr satisfies the latter assumption and
manages to capture the characteristic timescale over which source
intensities change. Moreover, we find that the chosen timestep allows
us to reliably track the morphological evolution of the reionisation
process during early epochs when the rate of sources switching on is
relatively fast.

Finally, we ignore the dynamical consequences of thermal feedback into
the gas from radiative ionisation, enabling us to decouple our
calculation of the radiation field from the hydrodynamical evolution
of the gas. This allows us to use existing outputs from cosmological
simulations to describe the evolving density field during the
reionisation process. In reality, photoheating introduces extra
thermal energy into the medium, and as ionisation fronts (I-fronts)
move from small scales to large scales, there is a corresponding
transfer of power from small to large scales through nonlinear
evolution. This effect is somewhat accounted for in the underlying
cosmological simulation used in this paper which includes a uniform
ionising background capable of heating the gas. However, one also
expects additional heating due to radiative transfer effects during
the reionisation process, and such uniform backgrounds cannot
reproduce the observed increase in gas temperatures from the extra
heating (Abel \& Haehnelt 1999 and references therein). As a result,
we expect minor systematic errors to be present in our solutions.
However, the main purpose of this paper is to describe the general
morphological evolution of the reionisation process on large scales,
which is insensitive to these systematic errors.  In particular,
recombination rates depend weakly on temperature and hence a proper
accounting of photons is possible even if the temperatures are not
computed precisely.

\subsection{Treating Multi-Species Reionisation}

One of the models we consider in this paper includes sources which are
capable of ionising both hydrogen and helium simultaneously.  In this
case, it becomes necessary to couple the corresponding ionisation
fractions during the radiative transfer calculations.  Exact solutions
require taking into account the detailed frequency dependences of all
the relevant processes and solving the multi-species chemical reaction
flows in nonequilibrium (e.g. Anninos et al. 1997). Owing to the large
number of sources which need to be processed in our study, the
implementation of a fully consistent nonequilibrium chemistry solver
would require computations that are not presently feasible.
Fortunately, there are a number of simplifying approximations which we
are able to employ that greatly speed up the calculations without
compromising the main objectives of our study.

Our analysis of hydrogen and helium reionisation requires us to track
three ionisation zones: H {\small II}, He {\small II}, and He {\small
III}.  In order to simplify our calculations, we decouple the
ionisation calculations by applying separate ray tracing calculations
for each species. To properly account for the recombination
rates implicit in the radiative transfer calculations, we are required
to follow a particular ordering for the ray tracing.  Specifically, since
He {\small III} ionisation zones will always reside within H {\small
II} zones, we first apply ray tracing calculations for He {\small II}
reionisation which incorporate recombination rates based on electron
densities that assume complete ionisation in both hydrogen and helium.
Although the equations for H {\small I} and He {\small II} ionisation
are, in principle, coupled, they can be separated to good accuracy
using the fact that recombination processes from He {\small III} to He
{\small II} are able to ionise H {\small I}.  In particular,
recombination emission from He {\small II} L$\alpha$ ($h\nu=40.7$ eV)
photons, which diffuse slowly owing to resonance scattering, and He
{\small II} Balmer-continuum photons, which are concentrated close to
the H {\small I} ionisation threshold, will tend to ionise H {\small
I} within the He {\small III} zone.  For near cosmic abundances, these
two processes are just about sufficient to balance H {\small II}
recombination within this zone (e.g. Osterbrock 1989). An additional
source of H {\small I} ionising radiation produced in this zone is He
{\small II} two-photon continuum emission for which $h\nu^{\prime} +
h\nu^{\prime\prime}=40.7$ eV (the spectrum peaks at 20.4 eV). Most of
these photons are able to escape out of the He {\small III} zone and
should, in principle, be added to to the pool of photons which will be
responsible for ionising the H {\small I} zone. In order to access the
maximal contribution from this process relative to the contribution
already present from the stellar radiation field, we examine the
hypothetical scenario where the IGM is completely doubly ionised in
helium by $z=10$.  Assuming a gas temperature of $2\times 10^4$ K and
a gas clumping factor of 30, we estimate that the total contribution
of ionising photons from He {\small II} two-photon emission will
comprise only 1\% of the total number of photons produced directly
from star-forming galaxies at the same redshift. Given the fact that
the uncertainties associated with intrinsic source intensities are
larger, we ignore this additional component to
simplify the calculations.

Once the He {\small III} zone has been determined, the next step
involves calculating the associated H {\small II} and He {\small II}
zones around the same source.  Here we note that the stellar radiation
field responsible for ionising the latter two species can only arise
from the hottest OB stars in the galaxies. This assumption is based on
the fact that only these stars have strong enough stellar winds
capable of producing {\it escape routes} through the host galaxy for
their ionising radiation.  Since the intrinsic spectra of OB stars
yield a large fraction of photons with $h\nu > 24.6$ eV, the outer
boundaries of the ionisation zones for H {\small II} and He {\small
II} coincide, thereby allowing the tracking of both zones via a single
ray tracing calculation. The approximation here will be to track H
{\small II} ionisations explicitly via ray tracing while implicitly
assuming that He {\small II} is ionised to the same degree and
extent. Since the relative cosmic abundance of helium is small ($\sim
8$\% by number density), the inaccuracies introduced by this
approximation should be negligible.  Note that in carrying out the
radiative transfer calculations for the ionisation of H {\small I}, we
are careful to ignore H {\small II} recombinations within the He
{\small III} regions for the reasons presented earlier.

After the ionisation zones have been determined, ionisation fractions
for He {\small III} and H {\small II} are calculated by applying the
following equilibrium condition for each case,
\begin{equation}
C_fn_en_{+}\alpha_{A}(T)=\Gamma n_o,
\end{equation}
where $C_f$ is the volume averaged gas clumping factor within the
cell, $\alpha_A(T)$ is the temperature dependent Case A recombination
coefficient, and $n_+$, $n_o$, and $n_e$ are the ionised, neutral, and
electron number densities, respectively. In this paper, we assume a
constant temperature of $2.0\times 10^4$ K for all regions which have
undergone ionisation to He {\small III} and $5.0\times 10^3$ K for
regions having undergone ionisation to H {\small II} and He {\small
II} only, (see Abel and Haehnelt 1999). This allows us to simplify our
calculations by using constant recombination coefficients.  From the
fitting formulae given in Hui \& Gnedin (1997), we compute the
relevant coefficients to be $\alpha_{A,\hehh}=1.40\times10^{-12}$
cm$^3$ s$^{-1}$ and $\alpha_{A,\hhh}=2.59\times10^{-13}$ cm$^3$
s$^{-1}$ for $T=2.0\times10^4$ K and
$\alpha_{A,\hh}=6.98\times10^{-13}$ cm$^3$ s$^{-1}$ for
$T=5.0\times10^3$ K. The photoionisation rate in an ionised cell for a
given species will be approximated by
\begin{equation}
\Gamma=\frac{\bar{\sigma}\dot{N}_{ph}}{4\pi r^2} \ \text{s}^{-1},
\end{equation}
where $\bar{\sigma}$ is the mean cross section for photoionisation,
$\dot{N}_{ph}$ is the number of relevant ionising photons released per
second from the source, and $r$ is the distance from the source to the
cell in question. Assuming that the sources have intrinsic spectral
energy distributions of the form $J(\nu)\propto\nu^{-\alpha}$ in the
relevant frequency range of interest and that
$\sigma(\nu)\propto\nu^{-3}$ beyond the ionisation edge, we can
approximate the mean photoionisation cross sections according to
\begin{equation}
\bar{\sigma}=\sigma_o\frac{\alpha}{\alpha+3},
\end{equation}
where $\sigma_o$ is the cross section at the ionisation frequency.
For stellar sources we choose $\alpha=1.8$ (up to 4 Ryd) based on the
extrapolated far-UV synthetic spectrum for starburst galaxies derived
in Haehnelt et al. (2001). Interestingly, this value is also consistent
with estimates for the average quasar spectral index (see the 
relevant discussion
in Sokasian et al. 2002), simplifying our analysis involving
reionisation by harder sources such as AGNs (see \S 5.4). In this
paper we adopt $\sigma_{\heh,o}=1.58 \times 10^{-18}$ cm$^2$ and
$\sigma_{\hh,o}=6.30 \times 10^{-18}$ cm$^2$ from Osterbrock (1989).

To further simplify our calculations, we remove the coupling between
the ionisation fractions of the two species by approximating separate
expressions for the electron densities in each ionisation zone.
Specifically, ionisation fractions for He {\small III} are computed
first by approximating the medium within the He {\small III} as fully
ionised in He {\small II} and H {\small I}, thus giving
\begin{equation}
n_e = (\chi_{\hehh}+1)n_{\phet} + n_{\pht},
\end{equation}
where $n_{\tiny{\mbox{\pht}}}$ and $n_{\tiny {\mbox{\phet}}}$
represent the total hydrogen and helium particle densities, and
$\chi_{\hehh}$ is the ionisation fraction for He {\small
III}.  Ionisation fractions for H {\small II} are then computed using
\begin{equation}
n_e = (\chi_{\hehh} +\chi_{\hhh})n_{\phet} + \chi_{\hhh}n_{\pht},
\end{equation}
where we have approximated $\chi_{\heh}=\chi_{\hhh}$ based on our
earlier assumptions related to the nature of the sources responsible
for ionising hydrogen and singly ionising helium. Again, since the
helium abundance is small the latter approximation should be
relatively accurate.  Although our procedure for calculating
ionisation fractions is not fully consistent, the resultant
inaccuracies are relatively small and certainly fall within the
broader uncertainties associated with the simulation as a whole.

The approximations listed above offer computational benefits that
allow us to process a large number of sources easily. Since they do
not compromise the main objective of this paper, which is to study the
morphological evolution of reionisation, we feel justified in adopting
them.

\section{UNDERLYING COSMOLOGICAL SIMULATION}

The cosmological simulation we will use in our analysis is based on a
SPH treatment of a $\Lambda$CDM model with parameters $\Omega_o=0.3$,
$\Omega_{\Lambda}=0.7$, Hubble constant $H_o=100\ h$ km s$^{-1}$
Mpc$^{-1}$ with $h=0.7$, baryon density $\Omega_b=0.04$, and a
scale-invariant primordial power spectrum with index n=1, normalised
to the abundance of rich galaxy clusters at the present day
($\sigma_8=0.9$). These choices are in good agreement with those
inferred from the WMAP observations (Bennett et al. 2003).

The simulation uses $324^{3}$ SPH particles and $324^{3}$ dark matter
particles in a $10.0 \ h^{-1}$ Mpc comoving box, resulting in mass
resolutions of $2.12\times 10^6\ h^{-1}$ M$_{\odot}$ and $3.26\times
10^5 \ h^{-1}$ M$_{\odot}$ in the dark matter and gas components,
respectively. This particular simulation represents one of the
high-resolution runs (simulation Q5) in the series of calculations
performed by Springel \& Hernquist (2003a).  These authors used a
large set of high-resolution simulations on interlocking scales and at
interlocking redshifts to infer the evolution of cosmic star formation
from high redshift to the present.  These simulations included a novel
description for star formation and feedback processes within the
interstellar medium and a treatment of galactic outflows (Springel \&
Hernquist, 2003b), and employed a conservative formulation of SPH
based on following the specific entropy as independent thermodynamic
variable (Springel \& Hernquist, 2002).The broad range of scales
encompassed by their set of simulations, together with extensive
convergence tests, enabled them to obtain a converged prediction for
the cosmic star formation rate, $\dot{\rho}_{\star}$, within the
adopted model for galaxy formation.

In the following section we will discuss the motivation behind our
particular choice for the simulation scale. We will also describe our
methodology for selecting sources and assigning intensities in the
context of the converged prediction for $\dot{\rho}_{\star}$ from
Springel \& Hernquist (2003a).

\section{SOURCE SELECTION}

To accurately simulate the reionisation process due to stellar
sources, one requires the cosmological volume to be sufficiently large
to properly sample a representative range of halo masses. At the same
time, one is inhibited from adopting an excessively large volume which
can compromise the technical feasibility of the simulation owing to the
large number of sources which would need to be processed. Furthermore,
one requires sufficient mass resolution within the simulated volume in
order to reliably measure star formation rates in galaxies. This is
especially true at high redshifts, where the bulk of the star
formation occurs in objects of relatively low mass, as expected for
hierarchical growth of structure, making it important to resolve
low-mass objects. From the set of simulation runs presented in
Springel \& Hernquist (2003a), we find that their Q5 run provides the
most appropriate compromise ($10.0 \ h^{-1}$ Mpc box length), capable of
capturing a representative sample of the halo population while
maintaining a level of mass resolution that reliably matches the
converged star formation rate.

\subsection{Source Definition}

Our approach for defining sources relies on identifying virialised
halos as sites of star formation. As in Springel \& Hernquist (2003a),
halos are located by employing a ``friends-of-friends'' (FOF) group
finder algorithm to the dark matter particles.  The algorithm is
restricted to dark matter particles only to avoid complications
associated with tracking baryonic particles whose number and type can
vary with time. The two main parameters which govern the group-finding
algorithm are the comoving linking length and the minimum number of
particles required for a group.  For the former, we choose a fixed
value equal to 0.2 times the mean interparticle spacing of dark mater
particles, corresponding to a group overdensity of roughly 200. The
choice for the minimum number of particles which can define a group is
somewhat more arbitrary.  This is due to the fact that this parameter
is invariably related to the question as to what exactly constitutes a
group. A reasonable definition may be that given a specific
overdensity, a group is considered to be {\it real} if it can be
reliablely tracked over time.  More specifically, this would
correspond to being able to find a given halo in a subsequent time
step. Below a certain threshold, one might start to identify {\it
transient} groups, that dissolve away in subsequent time steps. Our
statistics on halo tracking indicate that this threshold is around 100
dark matter particles, with associated uncertainty growing
progressively towards lower particle numbers; i.e. below about 100
particles one starts to occasionally pick up transient halos that are
probably flukes, and the fraction of these slowly grows towards smaller
groups. However, as we shall show in the following section, a
significant portion of the total star formation rate is predicted to
come from low-mass halos with fewer than 100 dark matter particles.
These objects essentially represent the star-forming mini-halos of the
Universe. In order to account for these sources, we adopt an
aggressively low minimum value of 16 particles for our definition of a
group.  This will inevitably raise the level of uncertainty regarding
the reality of some of the low-mass halos and their corresponding star
formation rates.  However, this approach provides us with a list of
low-mass objects whose number density and spatial locations are
statistically consistent with results from simulations conducted with
higher resolution. The star formation rates in these objects can then
be systematically corrected so that the total contribution from all
objects matches the converged result presented in Springel \&
Hernquist (2003a).

\subsection{Star Formation Rates}

Our prescription for assigning star formation rates to the groups
relies first on associating each gas or star particle with its nearest
dark matter particle, and the FOF-group this particle resides in. We
then take the total mass of each group as a ``viral'' mass, and the
sum of star formation rates in all of a group's SPH particles as the
total star formation rate in the halo. Our working assumption will 
then be that every halo hosts a single galaxy which will act as a
source. In reality, however, there may be more than one galaxy per
halo. The difference, however, is negligible in the context of the
spatial resolution of the radiative transfer calculations. More
specifically, in the redshift range of interest, even the most massive
halos have virial radii which are less than or comparable to the width
of a single cell.

In Figure 1, we plot various statistical quantities related to the
star formation rate from the resultant source list as a function of
halo mass for redshifts z = 14.7, 10.4, and 7.2.  At each redshift we
have binned the halos by mass in logarithmic intervals of dlog$M$ =
0.1 and computed the associated star formation rate density, number
density, fractional star formation contribution, and cumulative star
formation contribution below a given mass. From the first two rows of
plots, it is evident that while the relative amplitude in the star
formation rate becomes smaller towards the lower mass scales, the
number of halos rises dramatically.  Note also the overall rise of the
star formation rate with decreasing redshift. In Hernquist \& Springel
(2003), the authors used simple analytic reasoning to identify the
physical processes that govern this evolution. As part of their
conclusions, they find that at early times ($z>6$), densities are
sufficiently high and cooling times are sufficiently short that
abundant star-forming gas exists in all dark matter halos with 
$T_{\rm vir}>10^4\,{\rm K}$ that can cool by atomic line
cooling. Consequently, the evolution of the star formation rate is
dominated by gravitationally driven growth of the halo mass function,
as is seen here.  The relative contribution to the total cosmic star
formation rate density as a function of halo mass is shown in the
bottom two rows of Figure 1.  Here we see that intermediate mass
scales provide the largest contribution to the overall total, with a
progressively larger contribution from higher mass scales at
decreasing redshifts.

Owing to the finite dynamic range of the cosmological simulation,
we cannot expect to resolve {\it all} the star forming halos that
contribute to the ionising emissivity. This becomes progressively
problematic at high redshifts where most of the star formation takes
place in low-mass galaxies which form in abundant numbers as
progenitors of more massive systems. Furthermore, near the resolution
limit of identifiable halos, the corresponding star formation rate
becomes less reliable owing to the relatively small number of SPH
particles associated with the halo. The resultant star
formation missed by our simulation due to these resolution effects is
shown in Figure 2 for the redshift range of interest. In the top panel
we plot the total star formation rate from all the halos identified in
our simulation (dashed line) and compare it to the physically
motivated analytic fit to numerical converged results presented in
Springel \& Hernquist (2003a). The analytic fit, obtained from the
analysis of Hernquist \& Springel (2003), takes the form of
\begin{equation}
\dot{\rho}_{\star}=\dot{\rho}_{\star}(0)\frac{\chi^2}{1+\alpha(\chi-1)^3\exp(\beta
\chi^{7/4})},
\end{equation}
where,
\begin{equation}
\chi(z)=\biggl(\frac{H(z)}{H_o}\biggl)^{2/3},
\end{equation}
and $\alpha=0.012$, $\beta=0.041$, and $\dot{\rho}_{\star} =0.013$
M$_{\odot}$ yr$^{-1}$ Mpc$^{-1}$ define the fitting parameters.  The
above fitting function was motivated by the bimodal behaviour in the
evolution of the cosmic star formation rate. At low redshifts, the
efficiency of gas cooling processes is diminished owing to the expansion
of the Universe and hence manifests itself through a scaling that is
related to the expansion rate as quantified by the Hubble constant. At
high redshifts, the steep rise towards lower redshifts is
related to the growth of the halo mass function which exhibits an
exponential cut-off for large masses. To access the significance of
the discrepancy between the numerical and the theoretical results in
the
context of the reionisation of the Universe, we find it convenient to
relate star formation rates in terms of the cumulative number of
ionising photons released as a function of look-back time. In the
bottom panel of Figure 2 we plot this quantity in terms of a ratio
between the simulation results and the theoretical prediction.  Here
we have adopted a conversion factor of $10^{53}$ s$^{-1}$ ionising
photons per star forming rate of $1$ M$_{\odot}$ yr$^{-1}$ based in
the analysis from Madau et al. (1999). We find that in terms of the
cumulative number of ionising photons released, the discrepancy does
not become dramatic until one exceeds a redshift of $\sim 15$ or so.

\subsubsection{Correcting Star Formation Rates in Low-Mass Sources}

The discrepancy described above does not pose a serious obstacle in
the context of our present goals; we find that we can easily correct
for the missing star formation rate through a simple algorithm by
adjusting the rates in the low-mass halos.

We start first by deciding upon a mass threshold for halos above which
we are confident in our results for the associated star formation
rate.  After analysing the results from the group finding algorithm
applied to our simulation, we find that independent of redshift, halos
with masses $M > 3.0\times 10^8\ h^{-1}$ M$_{\odot}$ have
statistically reliable values for their star formation rates. Below
this threshold, a halo may simply not have enough SPH particles to
make it possible to 
reliably compute the star formation rate or the halo may be unresolved
altogether. The fact that our source list aggressively includes
low-mass objects allows us to partly compensate for the missing
objects. Although the {\it reality} of some fraction of these low-mass
objects may be questionable, the fact that the statistical properties
regarding their number densities and locations matches those from
higher resolution runs assures us that we can at least statistically
capture the effect from this population of sources.  By coupling the
criterion regarding the {\it reality} of low-mass object with the
criterion for confidence in the computed star formation rate, we can
distinguish, in technical terms, two types of problematic sources: (1)
objects whose {\it reality} is not questioned (n$_{\text{DM}}>100$),
but whose star formation rate may be unreliable ($M < 3.0\times 10^8\
h^{-1}$ M$_{\odot}$), and (2) objects whose {\it reality} and
corresponding star formation is questionable (n$_{\text{DM}}<100$ and
$M < 3.0\times 10^8\ h^{-1}$ M$_{\odot}$).  The net effect from both
these sets of sources is to cause a deficit in the total cosmic star
formation rate relative to the theoretical expectation.  Our method
for correcting this deficit involves supplementing all objects whose
masses fall below $3.0\times 10^8\ h^{-1}$ M$_{\odot}$, comprising
both sets of problematic sources, with additional amounts of star
formation in proportion to their mass. The total amount added to all
these objects at a given redshift is normalised so that it exactly
makes up the deficit relative to the theoretical prediction. In Figure
3 we plot the comoving number densities and the star formation rates
as a function of redshift for all the sources in the simulation. In
the context of the deficit in the cosmic star formation rate, we have
grouped the sources into three categories with the first column
representing sources which do not require any corrections and the
second and third columns including the aforementioned problematic
sources. Here the dashed line and solid line indicate the uncorrected
and subsequently corrected contribution to the total star formation
rate, respectively. It is evident from the plots that the corrections
to the low-mass sources do not appear to be drastic and we are
confident that the uncertainties related to these adjustments cannot
significantly affect our results.

Having adjusted star formation rates in the low-mass sources we can
now estimate the magnitude of the discrepancies in the total star
formation rate had we been limited to poorer mass resolutions. In
Figure 4 ({\it left panel}) we plot this discrepancy in terms of the
fraction of the total star formation contributed from sources with
halo masses above a certain mass.  The total star formation rate is
based on all halos above the resolution limit in our simulation
($M=3.9 \times 10^7\ h^{-1}$ M$_{\odot}$). In the same figure ({\it
right panel}) we also plot source number densities as a function of
redshift for the same mass ranges, with the total source density from
the simulation also shown ({\it solid line}). As in Figure 1, it is
evident that the subsequent fraction of star formation contributed by
relatively low-mass objects rises rapidly with redshift. The labeled
mass inequalities were chosen to roughly match the source mass
resolutions associated with recent numerical attempts to simulate IGM
reionisation by stellar sources (see Ciardi et al. 2003b and Razoumov
et al. 2002). Interestingly, we find that in the context of our
corrected source list, the quoted mass limits continue to
significantly underestimate the relative contribution to the total
star formation rate from low-mass objects even at intermediate
redshifts when the ionising emissivity in the IGM is expected to have
reached appreciable levels. In particular, at $z\simeq9.2$ the cases
with $M>1.0\times10^9\ h^{-1}$ M$_{\odot}$ and $M>0.3\times10^9\
h^{-1}$ M$_{\odot}$, which are close to the mass resolution of sources
in the 'M3' and 'S3' simulations described by Ciardi et al. (2003b),
appear to account for only $\simeq53\%$ and $\simeq81\%$ of the total
star formation rate, respectively, relative to our complete source
list.  It is of course possible to renormalise the star formation
rates in the massive (resolved) sources so that the total star
formation rate is corrected to agree with our converged
results. However, such a simple correction fails to address the
potentially significant impact the environments around sources can
have on the amount of ionising radiation that can escape into the IGM.
In particular, by concentrating the ionising flux away from low mass
system which are more likely to reside in less clumpy environments
than their more massive counterparts, one runs the risk of
systematically overestimating recombination rates.  We will explore
the ramifications from such corrections in context of our own results
in \S 5.1.

\subsection{Source Gridding}

Given the redshift range of interest, $6\lesssim z \lesssim20$, and
our time step for updating source lists, $\Delta t = 4.0 \times 10^7$
yrs, the total number of sources which require processing reaches
nearly $1.5 \times 10^5$. Since our radiative transfer calculations
are performed on a Cartesian grid with a limiting resolution, it is
logical to group together sources within cells as a method of reducing
the number of sources.  Owing to the high resolution of our grid
($200^3$), the overall number of sources is reduced only by
$2\%$. Such a marginal reduction in sources assures us that the
resolution of our grid provides excellent resolving power in regards
to the spatial distribution of the sources.

\section{RESULTS AND DISCUSSION}

In our analysis, we carefully follow the reionisation process from a
set of models parameterised by the escape fraction. In this context,
the escape fraction is defined to be the fraction of ionising photons
intrinsic to each source that escape and enter into the radiative
transfer calculations. It must be noted here that this parameter
inevitably also carries with it any uncertainties associated with the
underlying amplitude of the star formation rates computed for the
sources.  Nevertheless, our approach allows us to predict the epoch of
reionisation as a function of this parameter.  By comparing these
predictions with recent observational constraints on the escape
fractions and neutral densities near $z\sim6$, we are able to test the
overall consistency of the theoretical predictions of the computed
star formation rates. Along the way we are also able to make general
statements regarding the morphological aspects of reionisation.

\subsection{Global Ionisation Fractions}

The evolution of the global ionisation fraction in the simulation
volume provides a useful way of characterising the overall ionisation
state of the Universe as a function of redshift.  In Figures 5a and 5b
we show how the ionised mass-weighted and volume-weighted fraction,
respectively, evolve with redshift.  The rate of the evolution to a
completely ionised state is quite similar for all the models, although
there is a subtle trend towards a swallower rise of ionisation
fractions with decreasing escape fractions.  Nevertheless, in all
cases the reionisation epoch, which we define as the redshift when the
global volume-weighted ionisation fraction reaches $99\%$, occurs by
$z\sim 7$.  Later in this section we show how a more detailed analysis
involving statistical comparisons between simulated spectra and
observational results for flux transmission at $z\simeq6$ can lead to
more rigid constraints as to how early reionisation could have taken
place.

Figures 5a and 5b also show that the ionised mass-weighted fraction is
consistently larger than the volume-weighted fraction at redshifts
before reionisation.  This result is seen more clearly in Figure 5c
where we have plotted the corresponding ratio of the two fractions.
The larger values for the ionised mass-weighted fraction at early
redshifts is consistent with the general picture of the pre-overlap
epoch where sources are preferentially ionising their relatively dense
surroundings before affecting the more tenuous IGM. The fact that the
inequality between the mass and volume fraction is amplified with
decreasing escape fraction highlights an interesting morphological
feature related to the number and intensity of the sources responsible
for reionisation.  Namely, when escape fractions are low, a relatively
larger number of sources are directly involved in the process of
ionising a neutral density field.  Since the
ionisation zones associated with these sources are also smaller, the
net effect is that massive regions are more likely to be ionised from
the ionising radiation emanating from within rather than radiation
that has first traversed the less dense IGM.  This predilection
naturally manifests itself in relatively larger ratios for the
mass-to-volume ionisation fractions during the early stages of
reionisation in these cases. Eventually, intervening patches of
neutral material become ionised and there is a sharp rise in the
intensity of the ionising background as contributions from different
sources combine. As this diffuse component comes to dominate the
radiation field, all regions become uniformly ionised and the ratio of
mass-to-volume ionisation fraction converges to unity.

A visual illustration of the reionisation process is shown in Figure 6
where we plot a series of projected slices through the simulation
volume.  In each panel, a $0.25 h^{-1}$Mpc slice from the
$f_{esc}=0.10$ run is projected in both density and ionisation
fraction.  Source locations in each slice are denoted by white crosses
making it easier to follow how the ionised regions ({\it blue})
percolate to turn a neutral IGM ({\it yellow}) into one that is highly
ionised ({\it red}).  Compared with earlier studies of reionisation,
Figure 6 shows that morphologically the ionised regions in our
analysis trace the large scale distribution of sources.  In
particular, we do not find that reionisation occurs mainly from low-
to high-density regions, as would be the case if photons escaped
preferentially into the IGM before ionising the high density regions
near the sources.

The panels in Figure 6 graphically illustrate the large number of
sources that are involved in the highly inhomogeneous reionisation
process.  A more dramatic illustration of this point is shown in
Figure 7 where we have plotted a 3D image of the same simulation
showing iso-surfaces around the ionised regions at $z=12.08$.  This
highly inhomogeneous morphology represents a departure from the
general picture of reionisation proposed by Gnedin (2000)
where only a handful of bright sources are responsible for reionising
the Universe. In the following section we discuss how the morphology
of the reionisation process can affect the number of ionising photons
required to achieve full reionisation.

\subsection{Ionising Photon Budget}

The number of ionising photons per hydrogen atom required for the
ionised regions around sources to completely overlap has developed
into an interesting topic that is commonly addressed in studies of
this kind (Ciardi et al. 2003b, Razoumov et al. 2002, Haiman et
al. 2001, Miralda-Escud{\'e} et al. 2000, Gnedin 2000). To reliably 
estimate
this quantity with simulations requires a high level of resolution
capable of properly estimating gas clumping factors which govern the
overall recombination rate. This requirement is especially important
in light of the results presented in Haiman et al. (2001) which show
that mini-halos with temperatures below $\sim 10^4$K can dominate the
average clumping in the early stages of reionisation and significantly
increase the required photon budget necessary to achieve complete
overlap.  Given the high resolution inherent to our hydrodynamic
simulations and the corresponding radiative transfer grids, coupled
with our ability to probe clumping factors on sub-cell scales based on
the underlying distribution of SPH particles, we are able for the
first time to probe the scales needed to account for recombination
effects arising from such low-mass objects. This represents a
significant improvement upon previous attempts which have relied on
conservative assumptions related to the escape fraction to accommodate
unresolved clumping.

In Figure 8, we plot the cumulative number of escaping ionising
photons per hydrogen atom as a function of redshift.  The diamond
symbol on each curve corresponding to different values for $f_{esc}$
represents the point at which the volume ionisation fraction exceeds
$99\%$.  For $f_{esc}=0.10\ -\ 0.30$, between 2 and 3 ionising photons
are required to reionise the Universe while 4-5 ionising photons are
needed in the case where $f_{esc}=1.0$. The difference in the number
of required ionising photons resulting from variations in the escape
fraction arises from the interesting interplay between the evolution
of clumping factors and the mean density of the Universe. In
particular, the rapid evolution of the mean density plays a slightly
more important role than the evolution of clumping factors in the
redshift range of interest.  Therefore, a delay in reionisation to
lower redshifts requires fewer ionising photons since the decline in
gas densities outweighs the corresponding increase in clumping
factors, resulting in comparatively fewer recombinations occurring
than at higher redshifts.

Environmental factors influence the number of photons required to
produce reionisation. A related issue is whether the distribution of
ionising intensities among the sources themselves can also have an
important impact on this process. In particular, given the same
amplitude and evolution of the cosmic star formation rate, it is
interesting to explore how a redistribution of the ionising flux from
dim low-mass sources to more massive sources would affect
reionisation.  In Figure 9, we examine this scenario by way of
comparisons involving the volume-weighted ionisation fraction and the
cumulative number of ionising photons that are released.
Specifically, we examine two situations: 1) the case where we have
retained the full source list which includes sources down to our
resolution limit of $M\simeq3.9\times 10^7\ h^{-1}$ M$_{\odot}$, and
2) the case where we have systematically transferred ionising fluxes
from sources with $M<1.0\times 10^9\ h^{-1}$ M$_{\odot}$ to the
nearest neighbouring sources with masses above this limit. The figure
shows that the affect can be significant, namely that reionisation
takes place later and requires more photons (by roughly $\simeq 63\%$)
for the same escape fraction ($f_{esc}=0.20$) when low-mass sources
are excluded. This result stems from the fact that low-mass sources
preferentially reside in less clumpy environments than their massive
counterparts.  Consequently, the net effect of transferring ionising
flux from these sources to more massive systems amounts to increasing
the overall ionising intensity in regions with higher recombination
rates. This leads to recombinations playing a more dominant role and
thus delaying the onset of reionisation.  This finding therefore
supports the requirement that dim low-mass sources be properly
resolved and included in future simulations of this kind. In
particular, inaccuracies related to the inability to resolve low-mass
sources and their corresponding contribution to the cosmic star
formation rate may be amplified due to their preferential location in
less clumpy environments.

\subsection{Observational Comparisons: {\it Flux Transmittance}}
Recent discoveries of quasars at redshifts 5.8 and greater (Becker et
al. 2001, Fan et al. 2001, Fan et al. 2000) are finally making
possible quantitative studies of the status of the IGM at high
redshifts.  In Fan et al. (2000) the spectrum of a $z=5.8$ quasar
(SDSS J1010-0125) was examined and revealed no evidence for a
Gunn-Peterson trough, indicating that the IGM was highly ionised near
$z\sim5.5$. In subsequent observations (Fan et. al 2001) the authors
studied three new quasars at $z>5.8$ and found a significant increase
in Ly$\alpha$ absorption from redshift 5.5 to 6.0. In particular, they
found a $\sim 300$ \AA \ region in the Ly$\alpha$ forest portion of
the spectrum of a $z=6.28$ quasar (SDSS 1030.10+0524) which had a
measured flux transmittance consistent with zero, indicating a flux
decrement of $\gtrsim50$, and suggesting the possible detection of a
Gunn-Peterson trough.  To more accurately quantify these findings
Becker et al. (2001) obtained higher resolution spectra of the three
new $z>5.8$ quasars, allowing them to place better constraints on the
status of the high redshift IGM.

To relate our results to these observations, we extract artificial
Ly$\alpha$ absorption spectra from the simulation outputs and then
derive statistical probabilities for the resultant flux transmittance
which can readily be compared to the observational results.  We
extract artificial spectra using the particle information in the SPH
simulation coupled with the ionisation information from our radiative
transfer calculations. For each model, we generate 500 spectra along
randomly selected lines of sight (LOSs) between $z\simeq5.8-6.2$.  Our
procedure is similar to that in the TIPSY software package
(Katz \& Quinn 1995), but does not require that a LOS be perpendicular
to a box face in the simulation volume.  Each LOS has a unique and
arbitrary direction relative to the box coordinate system and wraps
through the simulation volume repeatedly via periodic boundaries.
Using the smoothing kernels of the SPH particles, gas densities and
temperatures are computed along a LOS at appropriately sampled
intervals.  The component of the peculiar velocity of the gas in the
direction of the LOS is also computed at each point.  Once all
physical quantities have been gathered, Voigt profiles are fitted to
each spectrum by interpolating between the corresponding
line-absorption coefficients provided in Harris (1948). It is
important to point out that in computing the line profiles we use a
minimum gas temperature of $5.0\times10^3$ K as a correction to the
SPH temperatures which exclude the extra heating introduced by
radiative transfer effects (see Abel and Haehnelt, 1999).

Our simulations were stopped at $z=5.8$, allowing us to make
comparisons with the observational results from the spectrum of the
$z=6.28$ quasar.  The presence of an apparent Gunn-Peterson trough in
the spectrum of this quasar makes it an especially interesting
case. By using the results of our simulated spectra to make
statistical comparisons with this region and another region with a
non-zero transmittance, we can constrain the evolution of the
reionisation process.  Our analysis is based on probability
distribution functions involving transmission levels that have been
averaged over the same redshift range as the observational
measurements.  Specifically, we focus on the redshift ranges
$5.95<z<6.16$ and $5.74<z<5.95$, corresponding to the regions over
which the average transmittance was measured by Becker et al. (2001)
in the spectrum of the $z=6.28$ quasar.  In Figure 10, we plot the
probability distribution functions of simulated transmissions averaged
over the two redshift ranges and overlay the corresponding $\pm 1
\sigma$ measurements from Becker et al. (2001) ({\it vertical-dashed
lines}).  The distributions were obtained using 500 LOS, although we
have only plotted a restricted range of transmissions for the purposes
of observational comparisons.

It is clear from Figure 10 that the ionising emissivities predicted in
the models with $f_{esc}=0.10 - 0.20$ provide the most plausible match
to the observational results in terms of transmission probabilities
for the two redshift ranges. For this choice, reionisation is expected
to have occurred near $z\simeq7.8$, although as the analysis here
demonstrates, the IGM remains fairly opaque until $z\sim6$.  This
spread in redshift between the time when the Universe first became
ionised and the time when opacities declined enough to allow non-zero
transmission in Ly$\alpha$ is important for the proper interpretation
of observed quasar spectra. In particular, it is incorrect to assume
that a volume-weighted ionisation fraction above $99\%$ will
necessarily produce non-zero transmission levels consistent with
current observations of quasar spectra at $z\simeq6$. A much more
careful analysis involving extraction of spectra from simulations
which include high precision information regarding ionisation
fractions is necessary in order to make meaningful comparisons with
observational results.

Interestingly, an escape fraction of $10-20\%$ is reasonably close to
the range of values ($f_{esc}\lesssim0.10$) that have been
observationally deduced from $z<3$ starburst galaxies (see Heckman et
al 2001; Hurwitz et al. 1997; Leitherer 1995). It is important to note
that these observational results may be underestimating the true
values of the escape fraction owing to undetected absorption from
interstellar components. Also unclear, in context of our analysis, is
the possible evolution of the escape fraction with redshift which may
lead to significantly different values at $z>6$. In any case, the fact
that we require a reasonable escape fraction in order to match the
observations lends credence to the overall consistency of the
theoretical predictions of the computed star formation rates.

\subsection{Hydrogen Reionisation By AGNs?}

Our radiative transfer calculations have been designed to handle the
simultaneous reionisation of both the hydrogen and helium components
of the IGM. This allows us to explore the scenario where objects with
spectra harder than stellar sources were primarily responsible for the
reionisation of the Universe. Such a scenario has been proposed by
e.g. Haiman \& Loeb (1998). In their analysis, a theoretical
extrapolation of the quasar luminosity function at fainter
luminosities and higher redshifts than currently detected appears to
favour reionisation by low-luminosity mini-quasars (or AGNs).
Reionisation by AGN type sources has also been supported on the
grounds that they may be the only type of sources capable of having
large enough escape fractions to allow enough ionising photons to
enter the IGM (see Wood \& Loeb 2000).

To explore the implications of sources with hard spectra reionising
the Universe, we have run a model in which the sources are capable of
ionising He {\small II} in addition to H {\small I}.  Specifically, He
{\small II} ionising rates are computed for the sources by assuming a
single power-law form for the spectrum, $f_{\nu}\propto
\nu^{-\alpha}$, normalised to deliver the same number of H {\small I}
ionising photons below 4 Ryd predicted from soft sources with
$f_{esc}=0.20$. This ensures that the hydrogen component of the IGM
will reach roughly the same level of ionisation by $z\sim6$ as in the
$f_{esc}=0.20$ stellar model and therefore will also provide a
reasonable match to the observational results of Becker et al. (2001;
see \S 5.3). Extrapolated results for the far-UV synthetic spectrum of
a stellar population with continuous star formation reveals a typical
quasar spectrum with $f_{\nu}\propto \nu^{-1.8}$ up to 4 Ryd. Adopting
the same spectral index above 4 Ryd for the hard sources allows us to
relate the rate of ionising photons, $\dot{N}_{ph}$, in He {\small II}
and H {\small I} directly with the expression
\begin{equation}
\dot{N}^{H I}_ph = \dot{N}^{He II}_{ph} \biggl(\biggl(\frac{\nu_{He
II}}{\nu_{H I}}\biggl)^{-1.8} - 1\biggl),
\end{equation}
where $(\nu_{He II}/\nu_{H I})$ is the ratio of ionisation threshold
frequencies for the two species. This leads to a conversion factor of
$\sim0.08$ He {\small II} ionising photons for every H {\small I}
ionising photon. 

In Figure 11, we show the resultant evolution in the volume-weighted
ionisation fraction for both soft and hard sources.  Note how the
hydrogen component in the IGM becomes reionised earlier in the case
where hard sources are invoked. This is due to the fact that the He
{\small III} regions now surrounding the sources act as an additional
source of H {\small I} ionising photons (see \S 2.1). The bottom panel
of Figure 11 shows that the helium component becomes reionised by
$z\simeq6.5$ in this case. Such an early reionisation epoch for helium
by high redshift sources such as AGNs should be testable in the coming
decade with the Next Generation Space Telescope. Nevertheless, an
interesting question we can explore currently is whether such a
scenario may also leave an imprint on He {\small II} ionisation levels
at lower redshifts where observations currently exist. In particular,
we would like to test whether a hypothetical population of AGNs which
suddenly turn off at $z=5$ will have lingering effects in terms of
subsequent ionisation levels when another known population of quasars
turn on.

For such an analysis, we re-ran our most realistic quasar model (model
5) from Sokasian et al. (2002) with the initial conditions that the
helium component in the IGM is doubly ionised everywhere in the
box. We present the results showing the subsequent evolution in the
volume-weighted ionisation fraction in Figure 12. Here the solid line
represents the scenario described above while the dotted line
indicates the original simulation where quasars turn on in a medium
that is only singly ionised in helium. In the case where the IGM is
initially fully ionised, we see that He {\small III} recombinations
quickly cause the ionisation fraction to decrease shortly after $z=5$
when the ionising emissivity from quasars is still low. Eventually,
the emissivity from quasars grows and the ionisation fraction starts
to rise again.  However by $z\simeq3$ enough recombinations have
already taken place since $z=5$ to effectively remove all traces of an
early reionisation epoch, leading to ionisation levels that are
governed only by the quasar population. It thus appears as if the
present observations cannot rule out the simple AGN scenario we have
considered here.

\subsection{Electron Optical Depth to the Surface of Last Scattering}

The recent tentative measurement of polarisation in the CMB by the
WMAP satellite (Kogut et al. 2003) suggests that reionisation occurred
in a more complex manner than indicated by the SDSS quasars alone
(e.g., Spergel et al. 2003). The measurements exhibit an excess in
the CMB TE cross-section spectrum on large angular scales ($\ell <7$)
yielding an electron optical depth to the CMB surface of last
scattering of $\tau_e=0.17$.  The uncertainty associated with this
measurement depends on fitting all parameters concerned with the TT
power spectrum and the TE cross power spectrum. Kogut et al. (2003)
obtain a 68\% confidence range of $0.13 < \tau_e < 0.21$ corresponding
to an instantaneous reionisation epoch which occurred between
$14<z<20$.  None of the models we have considered in this paper is
able reionise the Universe by such early redshifts.

The connection between the implications of the WMAP results and those
of the SDSS quasars for the evolution of the IGM is unclear.  In an
attempt to determine the extent to which the sources examined in
this paper would require additional ionising emissivity to provide a
consistent match with the WMAP measurements we consider an evolving
escape fraction of the form $f_{esc}(z)=0.20e^{k(z-6)}$. Here, a
positive value for the parameter $k$ has the effect of producing
ionisation levels consistent with the observations from the $z=6.28$
SDSS quasar, as in the model with $f_{esc}=0.20$, while leading to
relatively larger ionising emissivities at higher redshifts.  We note
that with this choice for $f_{esc}$, the ``escape fraction'' can
exceed unity.  In the context of our present analysis, this simply
means that the rate of production of ionising photons is higher than
would be implied for our choice of the star formation history and the
IMF.  For simplicity, we absorb these uncertainties in what follows
into our definition of the ``escape fraction.''

In Figure 13 we compare three different cases for this form: 1) a case
with a constant escape fraction of $f_{esc}=0.20$ ($k=0$; {\it
solid-line}), 2) a case with $k=0.13$ resulting in an evolving escape
fraction that rises to unity at $z\simeq18$ around when the first sources
turn on ({\it dotted-line}), and 3) a case with $k=0.33$ resulting in
an escape fraction that exceeds unity beyond $z\gtrsim11$({\it
dashed-line}). In panels b and c of Figure 13 we plot the resultant
evolution of the volume weighted H {\small II} ionisation fraction and
the evolution of the electron optical depth to electron scattering,
$\tau_e$, given by:
\begin{equation}
\tau_e(z)=\int_z^0\sigma_T n_e(z^{\prime})c \left|
\frac{dt}{dz^{\prime}} \right| dz^{\prime},
\end{equation}
where $\sigma_T=6.65\times10^{-25}$ cm$^2$ is the Thompson cross
section and $n_e(z^{\prime})$ is the mean electron number density at
$z^{\prime}$. In the calculation of $\tau_e$ we assume that helium
is singly ionised to the same degree as the hydrogen
component for $z>3$ and doubly ionised everywhere at $z<3$.

It is evident from Figure 13 that matching the the optical depth
measurement from WMAP with the sources studied here would require
substantially larger ionising emissivities at high redshifts.  In
particular, only the case with $k=0.33$, which reaches an escape
fraction greater that 10 near $z\simeq18$, is capable of producing the
measured electron optical depth. Such a large increase in the stellar
production rate of ionising photons would require either a greatly
enhanced star formation rate at high $z$ relative to that obtained by
Hernquist \& Springel (2003) or an IMF which becomes more top-heavy
with increasing redshift, as might be expected for population III type
stars.  In a recent paper Ciardi et al. (2003a) argue that a separate
population is not necessary if one adopts a top-heavy IMF for
population II type stars with an optimistically large value for the
escape fraction.  However, it is unclear whether the escape fraction
adopted in their analysis is meaningful because of their inability to
resolve either clumping on sub-grid scales or objects below $10^9$
M$_{\odot}$.  Furthermore, in their best-fit model, reionisation is
complete by $z\simeq13$. Such an early reionisation seems difficult to
reconcile with observations of the Gunn-Peterson effect measured in
the $z=6.28$ SDSS quasar (Becker et al 2001).

\section{CONCLUSION}

We have utilised a fast radiative transfer code to study the
reionisation of the Universe by stellar sources.  Our method has
been applied to a high resolution cosmological hydrodynamical
simulation which enables us to probe scales previously unresolved in
similar type of studies. By supplementing a {\it one-step} radiative
transfer code specifically designed for ionisation processes with an
adaptive ray-tracing algorithm, we are able to significantly speed up
the calculations and are able to handle a large number sources.

One of the main goals of this analysis was to study the effect of
low-mass sources residing in mini-halos on the overall morphology of
the reionisation process.  Analysis of numerically converged results
for star formation rates and halo mass functions in the underlying
cosmological simulation allowed us to assess the resolution effects
associated with low-mass objects and apply well motivated
corrections. With these corrections, we were able to reduce the
effective mass resolution limit for source objects to $M\sim4.0 \times
10^{7}\ h^{-1}$ M$_{\odot}$, which is roughly an order of magnitude
smaller than previous studies of this kind. Our calculations reveal
that the process by which ionised regions in the IGM percolate is
highly inhomogeneous and especially sensitive to the inclusion of dim
sources.  More specifically, we find that given the same level of
cosmic star formation, the number of ionising photons required to
reionise the Universe is overestimated by $\sim63\%$ when only objects
with masses $>1.0\times10^9 \ h^{-1}$ M$_{\odot}$ are used as source
locations. This result stems from the fact that low-mass sources
preferentially reside in less clumpy environments than their massive
counterparts.  Consequently, their exclusion has the net effect of
concentrating more of the cosmic ionising radiation in regions which
have higher recombination rates.

Our results from the full source list that includes objects down to
$M\sim4.0 \times 10^{7}\ h^{-1}$ M$_{\odot}$ appear to show a similar
evolution to reionisation for various values of the escape fraction.
In all cases, the epoch of reionisation, which we define as occurring
when the global volume-weighted ionisation fraction reaches $99\%$,
happens by $z\sim 7$.  This result is consistent with the absence of
a Gunn-Peterson trough in observations of quasars below $z\sim6$. To
relate our results to these observations, we extract artificial
Ly$\alpha$ absorption spectra from simulation outputs and then derive
statistical probabilities for the resultant flux transmittance.  Our
comparisons are based on probability distribution functions involving
transmission levels that have been averaged over the same redshift
range as the observational measurements.  Specifically, we focus on
the redshift ranges $5.95<z<6.16$ and $5.74<z<5.95$, corresponding to
the regions over which the average transmittance was measured by
Becker et al. (2001) in the spectrum of their $z=6.28$ quasar.  We
find that given the amplitude and form of the underlying star
formation predictions, an escape fraction near $f_{esc}=0.20$ is most
consistent with the observational results. This leads to a
reionisation epoch that is complete by $z\simeq 7.8$ and requires
roughly $\sim3$ ionising photons per hydrogen atom.

In an attempt to explore the implications arising from a scenario where
sources with harder spectra such as AGNs are responsible for hydrogen
reionisation, we have analysed a model in which the spectrum of
sources is allowed to extend beyond 4 Ryd. As a result, sources are
able to reionise He {\small II} simultaneously with H {\small I}. Our
analysis shows that helium reionisation occurs near $z\simeq 6.5$ when
the spectrum is normalised to deliver the same number of H {\small I}
ionising photons below 4 Ryd as predicted from soft sources with
$f_{esc}=0.20$.  In a hypothetical case where these hard sources
abruptly turn off before the onset of quasars below $z=5$, we find
that an early reionisation of the helium component is not necessarily
in conflict with observational constraints of He {\small II} opacities
obtained at $z\simeq3$. This is due to the fact that by $z\simeq3$,
enough He {\small III} recombinations haven taken place to cause
ionisation levels to converge to the values governed by quasar
emissivities alone.

In this paper, we have focused on reionisation by stellar sources
similar to those seen in local galaxies; i.e. population II type
stars.  Our results show that the star-formation history inferred by
Springel \& Hernquist (2003a) based on detailed hydrodynamic
simulations can account for the properties of the IGM measured from
quasars at $z\simgt 6$.  To the extent that these comparisons are
confirmed by future theoretical and observational studies, our results
support the evolutionary history of ``ordinary'' star formation
derived analytically by Hernquist \& Springel (2003).

The recent measurements by the WMAP satellite (Kogut et al. 2003)
indicate that a large fraction of the IGM may have been ionised early,
by $z\sim 14-20$.  Taken together with the observations of the SDSS
quasars at $z\sim 6$, the implications of this result for the
evolution of the IGM are unclear.  Cen (2002) has proposed a scenario
in which the Universe was reionised early by population III stars, but
then much of the IGM recombined once these stars were no longer able
to form, and the Universe was reionised a second time by the next
generation of stars.  Our results in this paper with a constant escape
fraction would describe the second stage of reionisation in this
model, provided that the IGM mostly recombined at intermediate
redshifts.

Our models with a constant escape fraction predict a relatively late
reionisation epoch and cannot account for the optical depth inferred
by WMAP.  In an effort to study how this difference could be
reconciled in the context of population II type stars, we employed
models with evolving escape fractions and found that if $f_{esc}$ were
to rise from $f_{esc}=0.20$ near $z\simeq6$ to $f_{esc}\gtrsim10$ near
$z\sim18$, the WMAP and SDSS constraints can both be satisfied.  In
this picture, an ``escape fraction'' larger than unity implies that
the stellar production rate of ionising photons is much higher than
that which would be produced according to our model assumptions.  One
mechanism that would boost the production rate of ionising photons
would be if the IMF evolved so that it became more and more top-heavy with
increasing redshift.

In the future, we will combine simulations of early structure
formation with a detailed treatment of radiative transfer to
investigate the combined effects of reionisation by both population
III and population II stars.  Preliminary work on related issues by
Yoshida et al. (2003) indicates that dynamical heating by ongoing
accretion limits the rate at which cold, dense clouds of molecular
material can form in halos.  In principle, this effect will severely
reduce the rate of formation of massive stars through molecular
cooling, perhaps rendering them unimportant as sources of ionising
radiation.  Dynamical simulations will be required to understand these
processes in detail, and to reliably calibrate semi-analytical
treatments, in a manner similar to the approach described here.

\begin{acknowledgments}

We thank Matias Zaldarriaga for discussions regarding the statistical
significance of observational results in the context of our analysis.
We also thank Brant Robertson for his services as a system
administrator for the computer cluster used in this analysis.
A.S. thanks Daniel Harvey for many constructive discussions related to
this study.  This work was supported in part by NSF grants ACI
96-19019, AST 98-02568, AST 99-00877, and AST 00-71019 and NASA ATP
grant NAG5-12140.  The simulations were performed at the Center for
Parallel Astrophysical Computing at the Harvard-Smithsonian Center for
Astrophysics.

\end{acknowledgments}

\begin{acknowledgments}

\end{acknowledgments}

\clearpage
\begin{figure}[htb]
\figurenum{1}
\setlength{\unitlength}{1in}
\begin{picture}(6,6.5)
\put(-0.45,-1.9){\includegraphics{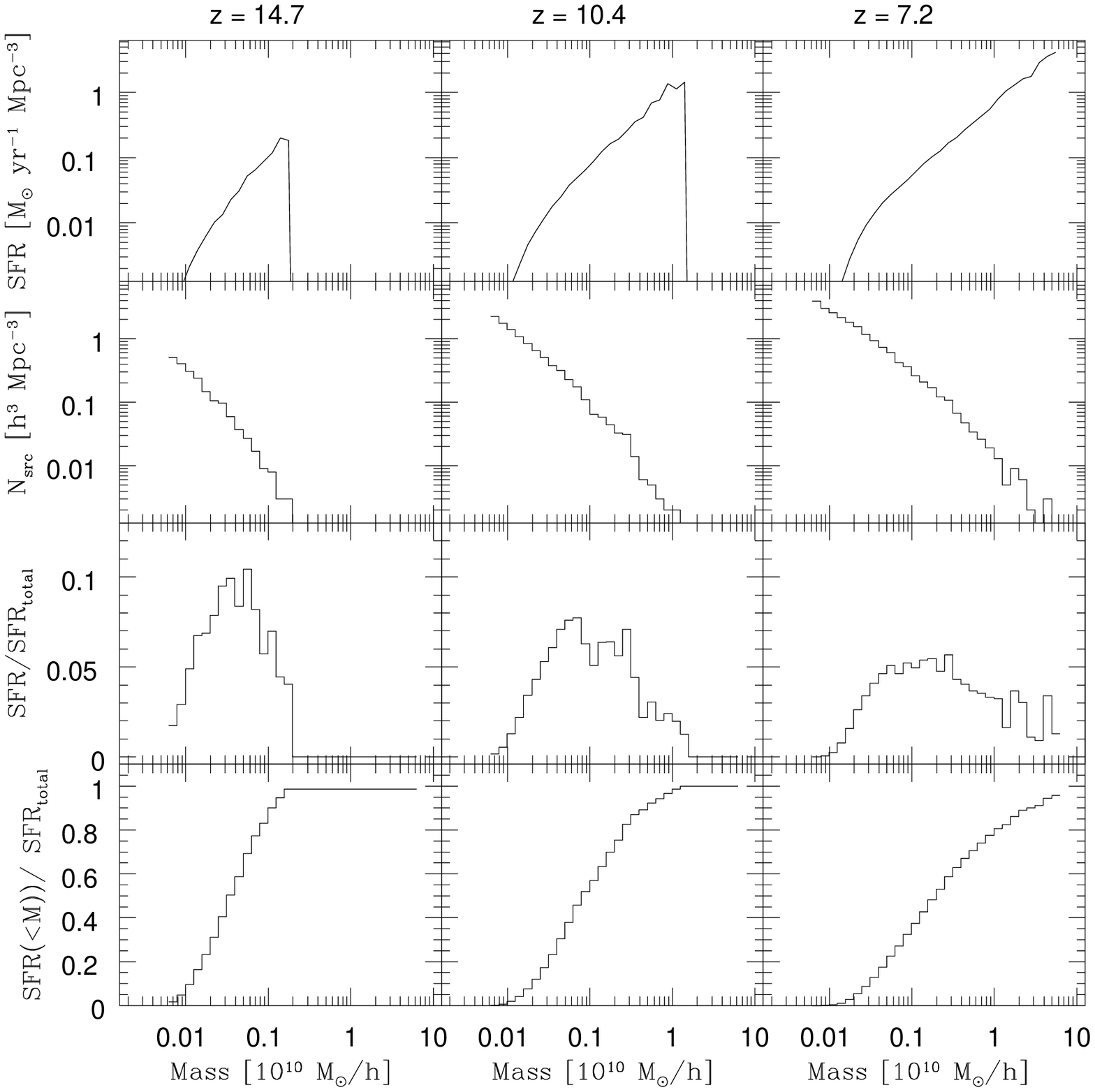}}
\end{picture}
\caption{Statistics related to our raw source lists for redshifts z=
14.7, 10.4 and 7.2.  The first row of panels show the star formation
rate density as a function of total halo mass, demonstrating how star
formation shifts to progressively larger mass scales with time. In the
second row, we plot the comoving number density of halos in
logarithmically spaced mass bins. The third row shows the fractional
contribution to the total cosmic star formation rate at a given time
for each mass bin. Note how at early times the contribution is heavily
weighted towards the lower mass scales. Finally in the fourth row we
plot the the cumulative fractional contribution to the total star
formation from all halos below mass M.}
\end{figure}

\clearpage
\begin{figure}[htb]
\figurenum{2}
\setlength{\unitlength}{1in}
\begin{picture}(6,6.5)
\put(1.07,-1.9){\includegraphics{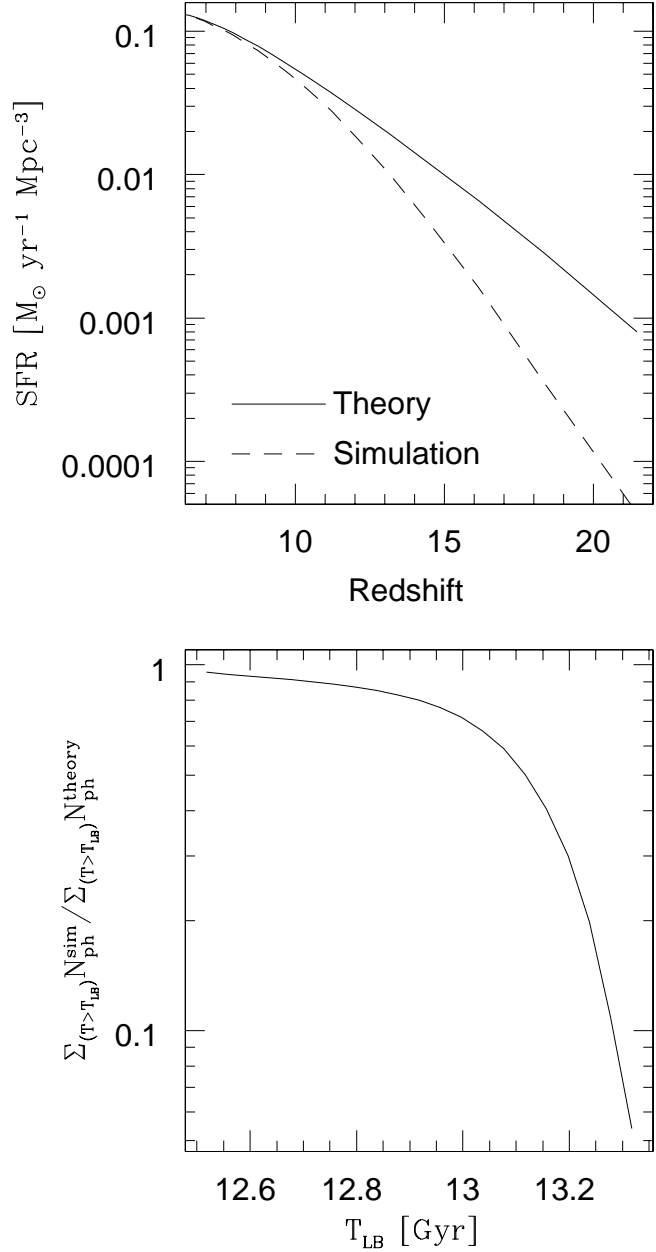}}
\end{picture}
\caption{Comparisons between the computed simulation results for the
cosmic star formation rate evolution and the theoretical prediction
from the fit to the converged results from Springel \& Hernquist
(2003a) obtained by Hernquist \& Springel (2003). In the top panel,
we show the evolution in star formation
rate as function of redshift for the two cases. In the bottom panel,
we plot the ratio of the simulation result to the theoretical
prediction in terms of the cumulative number of ionising photons
released as a function of look-back time.}
\end{figure}

\clearpage
\begin{figure}[htb]
\figurenum{3}
\setlength{\unitlength}{1in}
\begin{picture}(6,5.5)
\put(-0.67,-3.4){\includegraphics{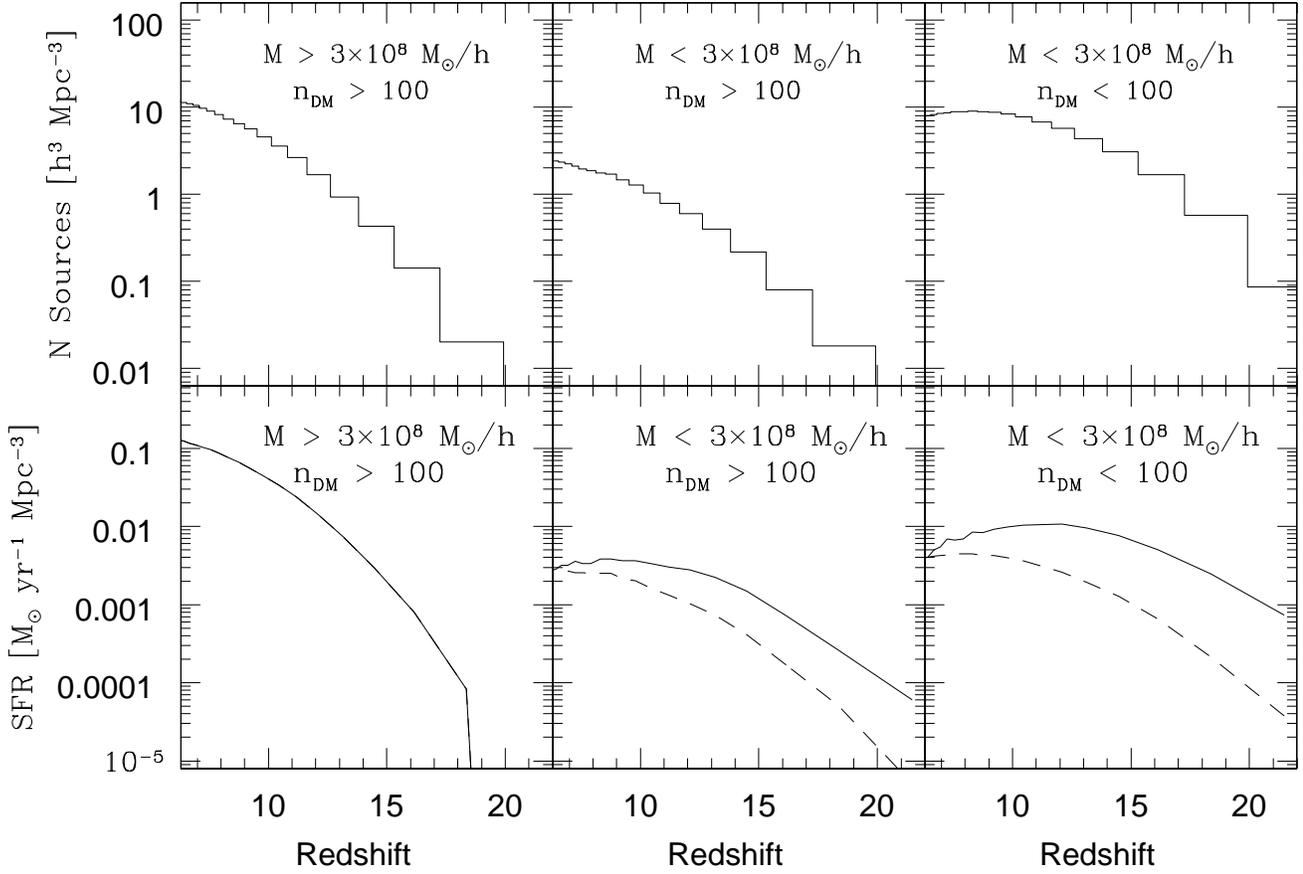}}
\end{picture}
\caption{To study the resolution effects responsible for the deficit
in the cosmic star formation rate relative to theoretical predictions,
we have grouped the sources according to the number of dark particles
and total mass in the associated halo. The top row of panels show
source comoving number densities at each redshift bin. In the bottom
row of panels we show the uncorrected ({\it dashed}) and corrected
({\it solid}) contributions to the cosmic star formation rate. The
first column represents massive sources which are not
affected by the limited resolution of the simulation. In the second
column we have sources whose {\it reality} is not questioned
(n$_{\text{DM}} > 100$), but whose associated star formation rate may
be underestimated due to the paucity of SPH 
particles associated with halos
with $M < 3.0\times10^8 \ h^{-1}$ M$_{\odot}$. And, finally, in the
third column we have the case where both the halo's existence is
questionable and the associated star formation rate is
underestimated. In the latter two cases, the star formation rate for
each source was increased in proportion to the corresponding halo mass
so that the total amount added equaled the deficit in the cosmic star
formation rate (see text for discussion).}
\end{figure}

\clearpage
\begin{figure}[htb]
\figurenum{4}
\setlength{\unitlength}{1in}
\begin{picture}(6,5.5)
\put(-.57,-3.9){\includegraphics{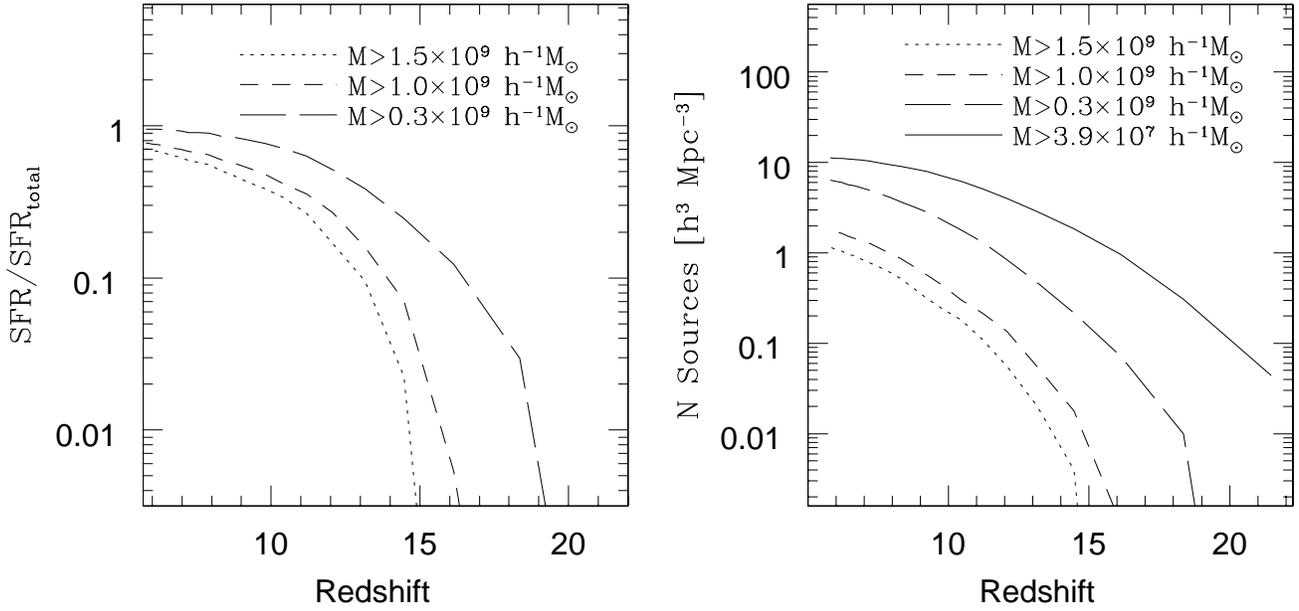}}
\end{picture}
\caption{{\it Left panel:} Evolution of the fractional star formation
rate when only objects above the labeled masses are included. The
total SFR is taken from our own corrected source list which includes
objects down to $M\simeq3.9\times 10^7 \ h^{-1}$ M$_{\odot}$. {\it
Right panel:} Comoving number densities of sources as a function of
redshift for the same mass ranges. The total number density from our
complete list of sources is also shown ({\it solid line}).}
\end{figure}

\clearpage
\begin{figure}[htb]
\figurenum{5}
\setlength{\unitlength}{1in}
\begin{picture}(6,5.5)
\put(1.00,-2.2){\includegraphics{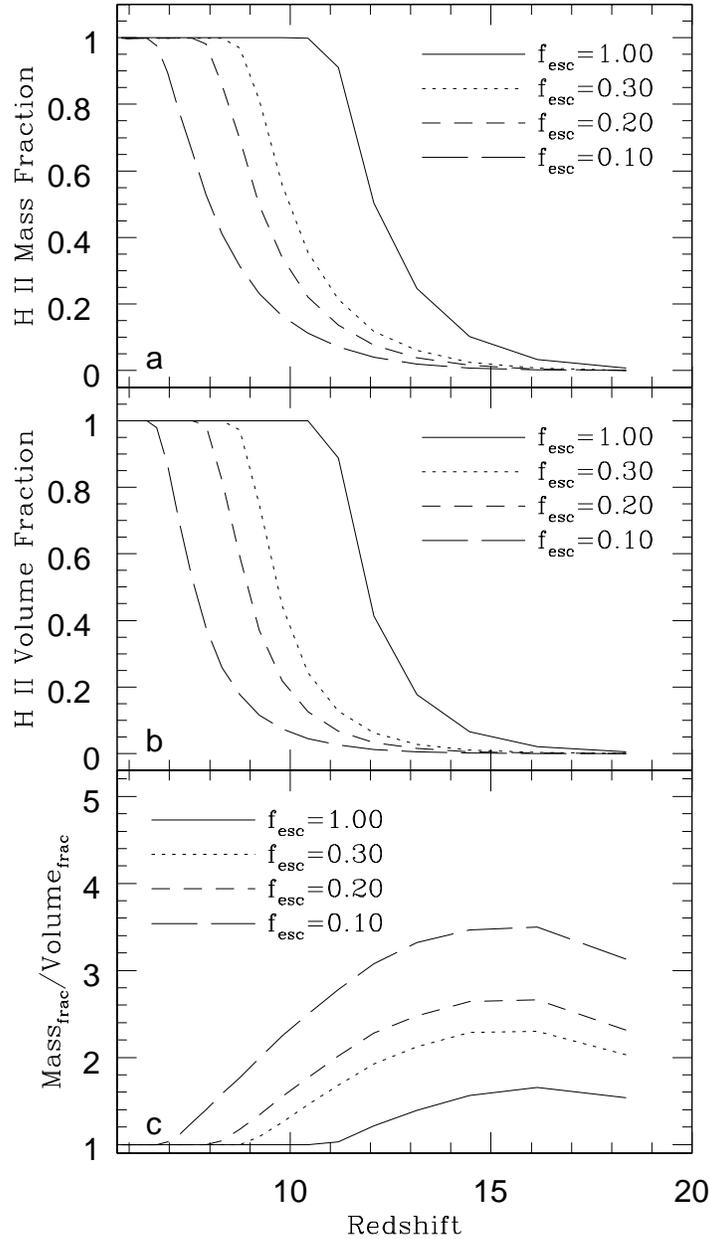}}
\end{picture}
\caption{Evolution of the ionised mass-weighted fraction (a), ionised
volume-weighted fraction (b), and the ionised mass to volume ratio (c)
as a function of redshift for the labeled values of escape fractions.
The epoch of reionisation is reached later with decreasing values of
$f_{esc}$ although the form of the evolution remains fairly similar in
each case. Note also that the mass-weighted fraction is consistently
greater than the volume-weighted fraction especially at early times
when much of the ionising emissivity is preferentially concentrated
in massive regions surrounding the sources.}
\end{figure}

\clearpage
\begin{figure}[htb]
\figurenum{6}
\setlength{\unitlength}{1in}
\begin{picture}(6,6.5)
\put(-0.45,-0.7){\includegraphics{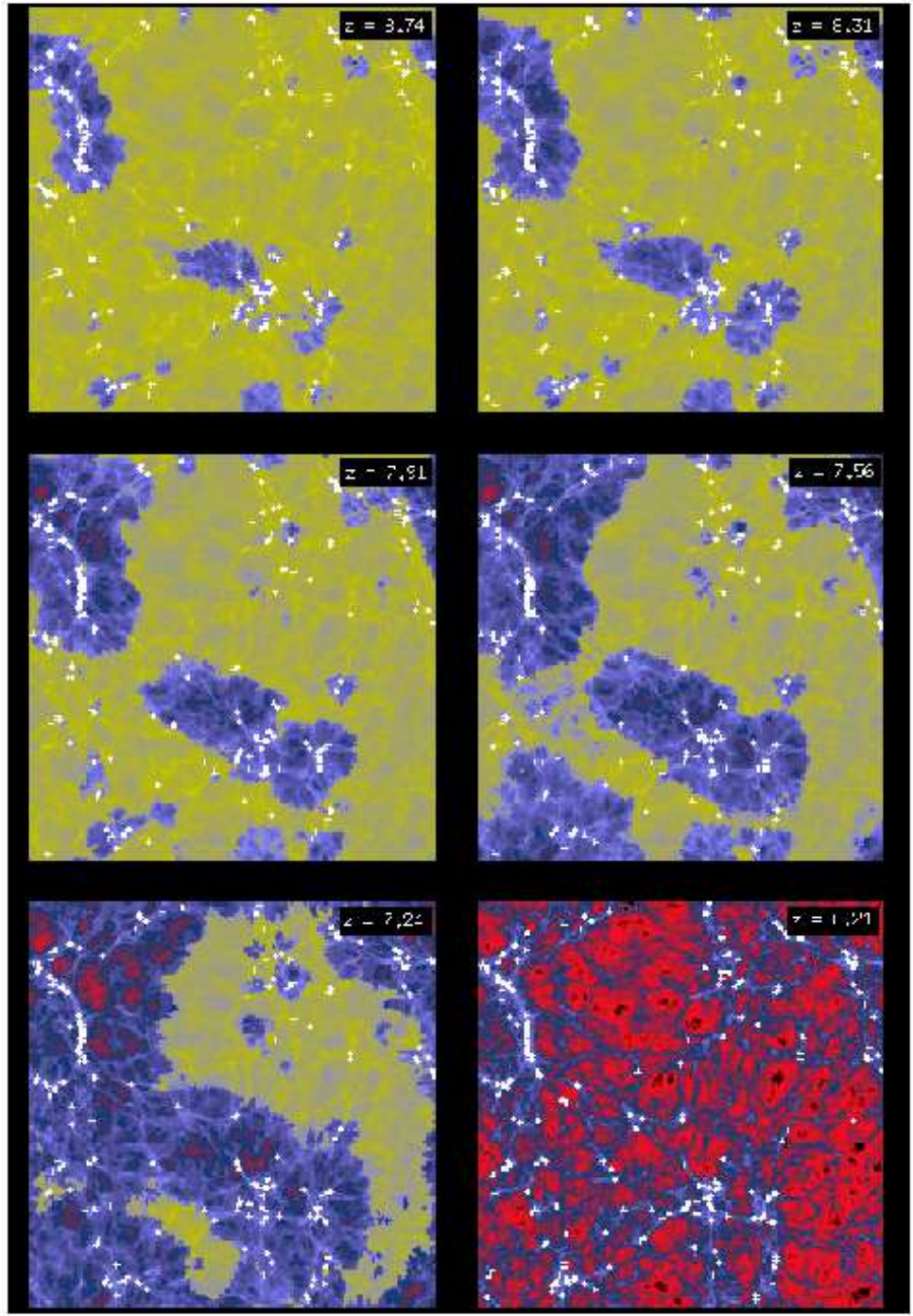}}
\end{picture}
\caption{A series of projected slices through the simulation volume at
({\it top-left} to {\it bottom-right}) z=8.74, 8.31, 7.91, 7.56, 7.24,
and 6.21. In each panel, a $0.25 h^{-1}$Mpc slice (1/40th of
a box length)
from the outputs of the $f_{esc}=0.10$ run is projected in both
density and ionisation fraction.  Source locations in each slice are
denoted by white crosses making it easier to follow how the ionised
regions ({\it blue}) percolate to turn a neutral IGM ({\it yellow})
into one that is highly ionised ({\it red}).}
\end{figure}

\clearpage
\begin{figure}[htb]
\figurenum{7}
\setlength{\unitlength}{1in}
\begin{picture}(6,6.5)
\put(-0.45,-0.9){\includegraphics{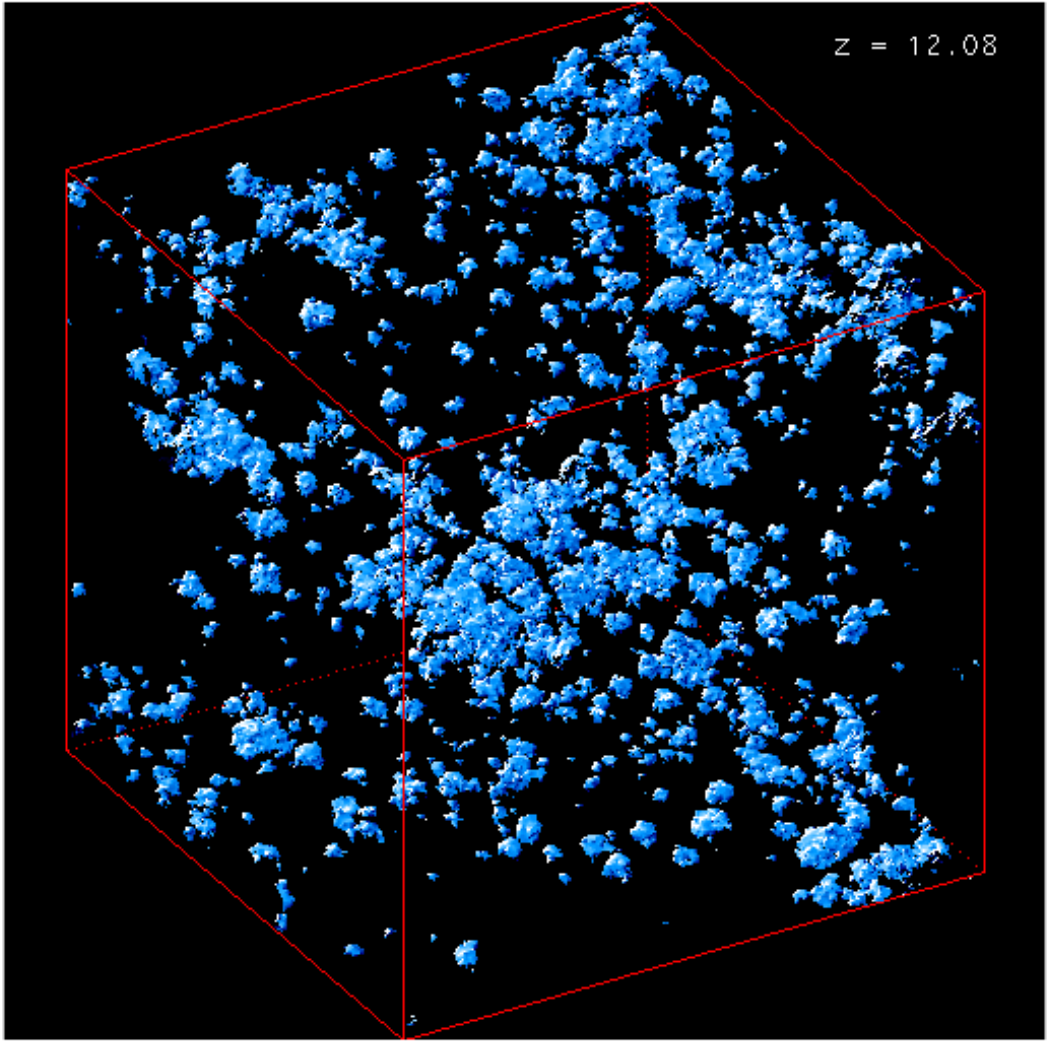}}
\end{picture}
\caption{A 3D image showing iso-surfaces around ionised regions
($\chi > 98\%$) at $z=12.08$ from the $f_{esc}=0.10$ model. The highly
inhomogeneous morphology associated with how sources ionise their
surroundings has a significant impact on the overall evolution of the
reionisation process (see text for discussion).}
\end{figure}

\clearpage
\begin{figure}[htb]
\figurenum{8}
\setlength{\unitlength}{1in}
\begin{picture}(6,5.5)
\put(1.0,-4.9){\includegraphics{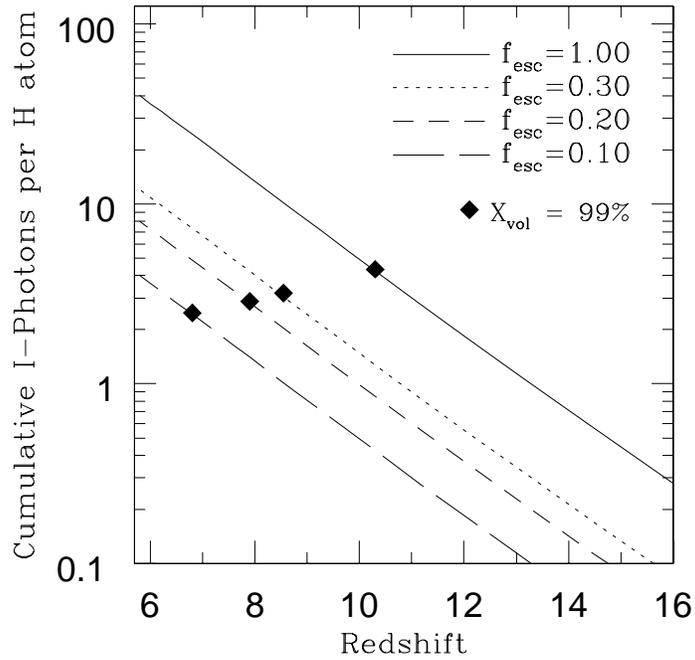}}
\end{picture}
\caption{Cumulative number of escaping ionising photons per hydrogen
atom as a function of redshift. The diamond symbol on each corresponding
case of $f_{esc}$ represents the point at which the volume-weighted
ionisation fraction exceeds $99\%$, which is our criterion for the
completion of the overlap epoch.}
\end{figure}

\clearpage
\begin{figure}[htb]
\figurenum{9}
\setlength{\unitlength}{1in}
\begin{picture}(6,5.5)
\put(1.1,-1.7){\includegraphics{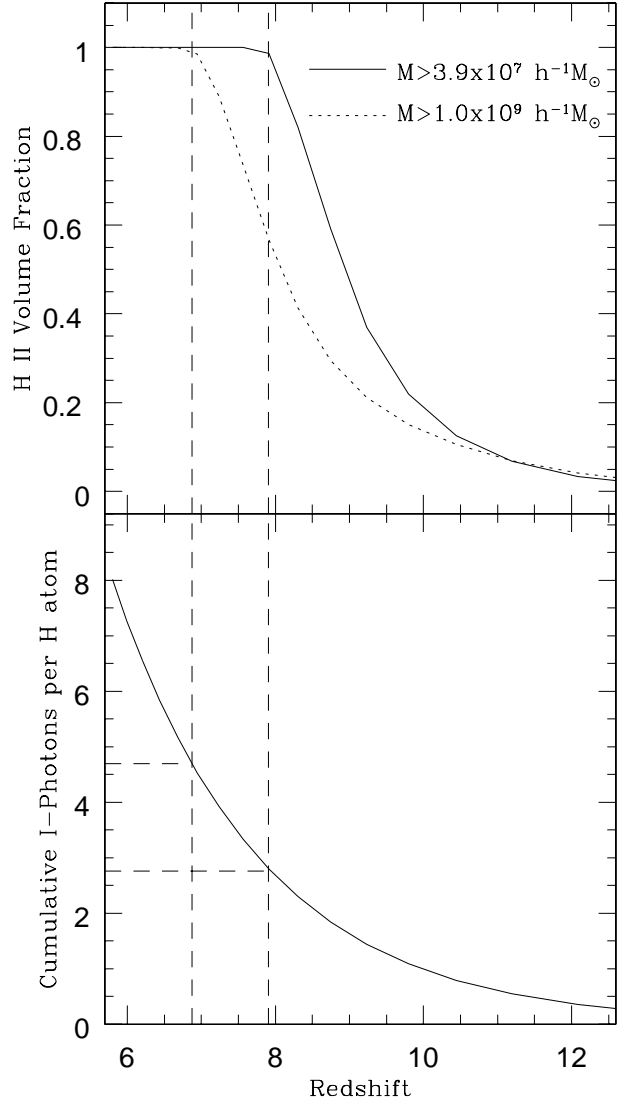}}
\end{picture}
\caption{{\it Top panel:} Evolution of the volume-weighted ionisation
fraction for the case where we have retained the full source list
which includes sources down to our resolution limit of
$M\simeq3.9\times 10^7\ h^{-1}$ M$_{\odot}$ ({\it solid line}), and
the case where we have systematically transferred ionising flux from
sources below $1.0\times 10^9 \ h^{-1}$ M$_{\odot}$ to the nearest
neighbouring sources with masses above this limit ({\it
dotted-line}). In both cases the escape fraction was set to
{$f_{esc}=0.20$}. {\it Bottom-panel:} Corresponding number of ionising
photons released as a function of redshift. Note that the completion of
overlap ({\it dashed-lines}) requires more ionising photons in the
case where ionising flux is transferred away from low-mass sources
(see text for discussion).}
\end{figure}

\clearpage
\begin{figure}[htb]
\figurenum{10}
\setlength{\unitlength}{1in}
\begin{picture}(6,6.4)
\put(-0.45,-1.8){\includegraphics{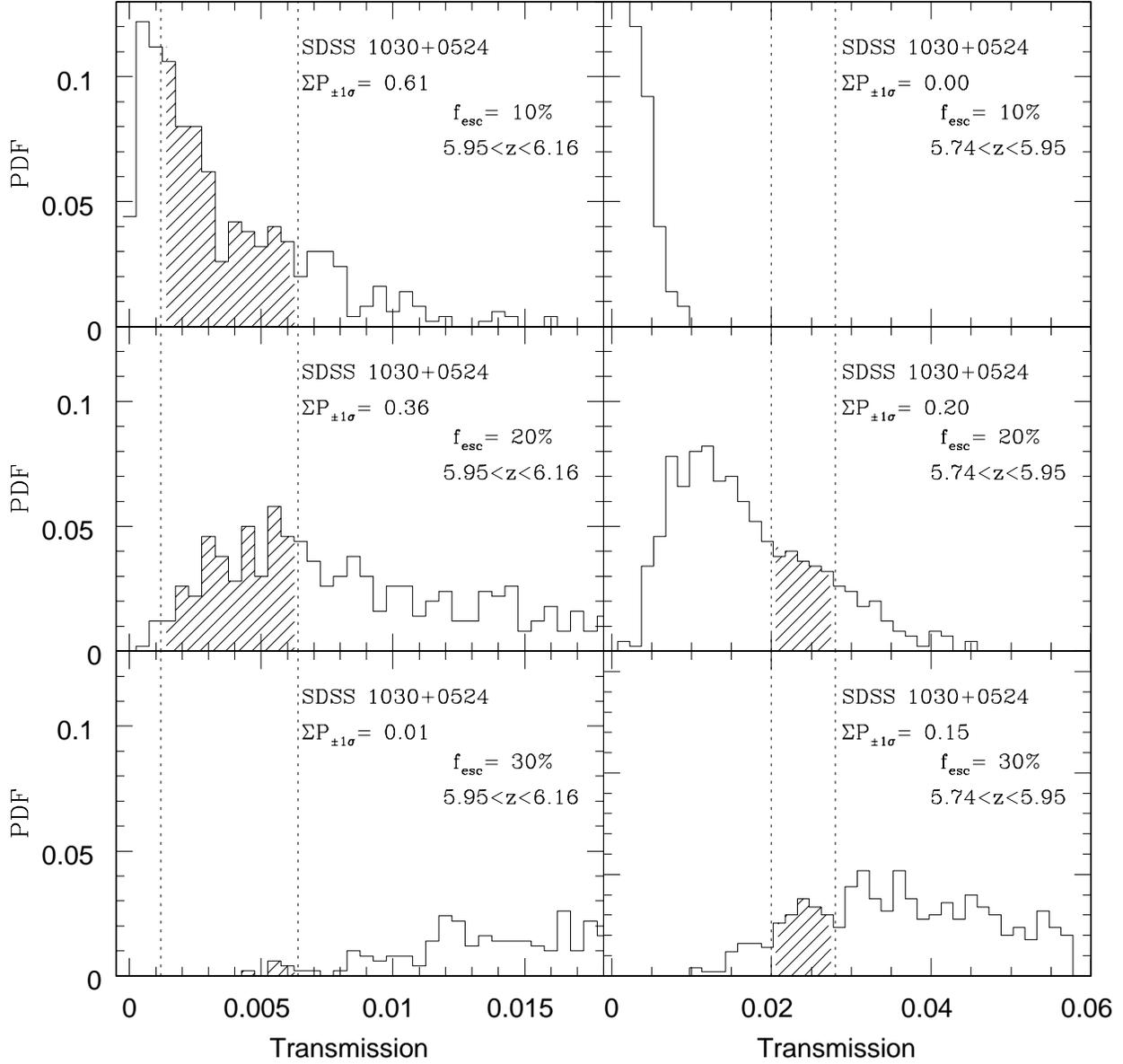}}
\end{picture}
\caption{Probability distribution functions of the average
transmission measured in the redshift ranges $5.95<z<6.16$ ({\it
left-column}) and $5.74<z<5.95$ ({\it right-column}) computed using
500 LOS from our simulations for $f_{esc}=0.10$, 0.20, and 0.30.  In
each column, the range of transmissions shown is chosen on the basis
of comparisons with the $\pm 1 \sigma$ measurements
reported by Becker et al. (2001) for the $z=6.28$ quasar ({\it
vertical-dotted lines}).  The cumulative probability of measuring
simulated transmissions consistent with the observations is
represented by the shaded regions and numerically labeled in each
panel.}
\end{figure}

\clearpage
\begin{figure}[htb]
\figurenum{11}
\setlength{\unitlength}{1in}
\begin{picture}(6,5.5)
\put(0.7,-3.9){\includegraphics{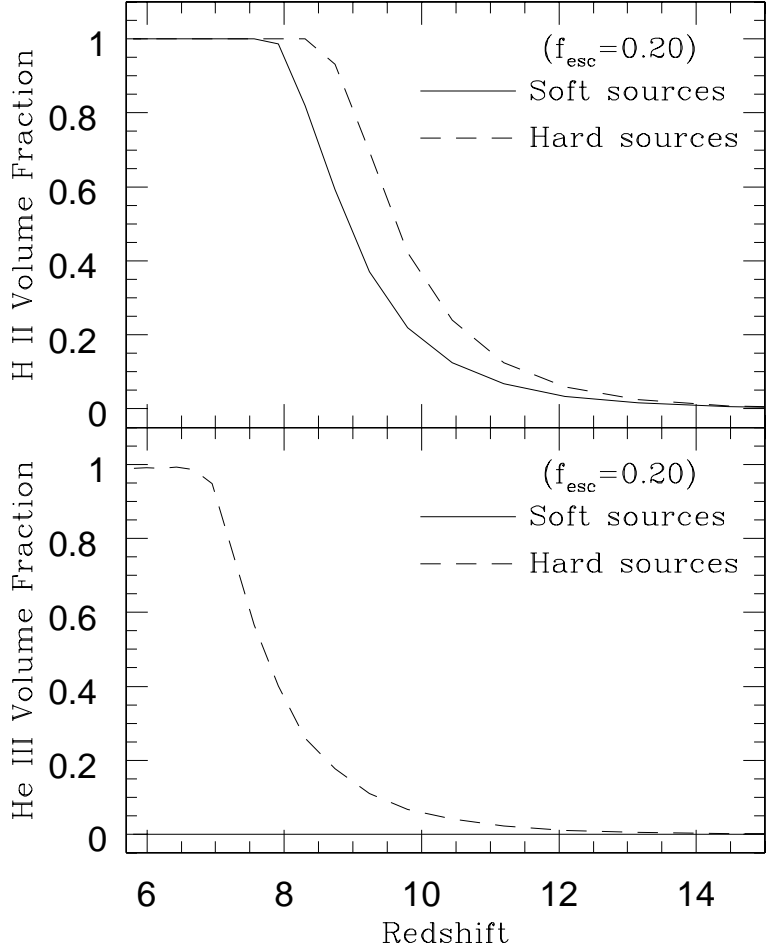}}
\end{picture}
\caption{Evolution of the H {\small II} ({\it top}) and He {\small
III} ({\it bottom}) volume-weighted ionisation fraction resulting from
sources with a soft spectrum capable of ionising only hydrogen
(stellar source model with $f_{esc}=0.20$; {\it solid-line}) and hard
sources capable of ionising both hydrogen and helium (AGNs; {\it
dashed-line}). He {\small II} ionising rates for the hard sources were
computed assuming a spectrum of the form, $f_{\nu}\propto \nu^{-1.8}$,
which is normalised to deliver the same number of H {\small I}
ionising photons below 4 Ryd as predicted from soft sources with
$f_{esc}=0.20$. Hydrogen reionisation occurs earlier in the case where
hard sources are invoked due to the additional H {\small I} ionising
emissivity introduced from He {\small II} reionisation.}
\end{figure}

\clearpage
\begin{figure}[htb]
\figurenum{12}
\setlength{\unitlength}{1in}
\begin{picture}(6,6.4)
\put(-0.45,-1.8){\includegraphics{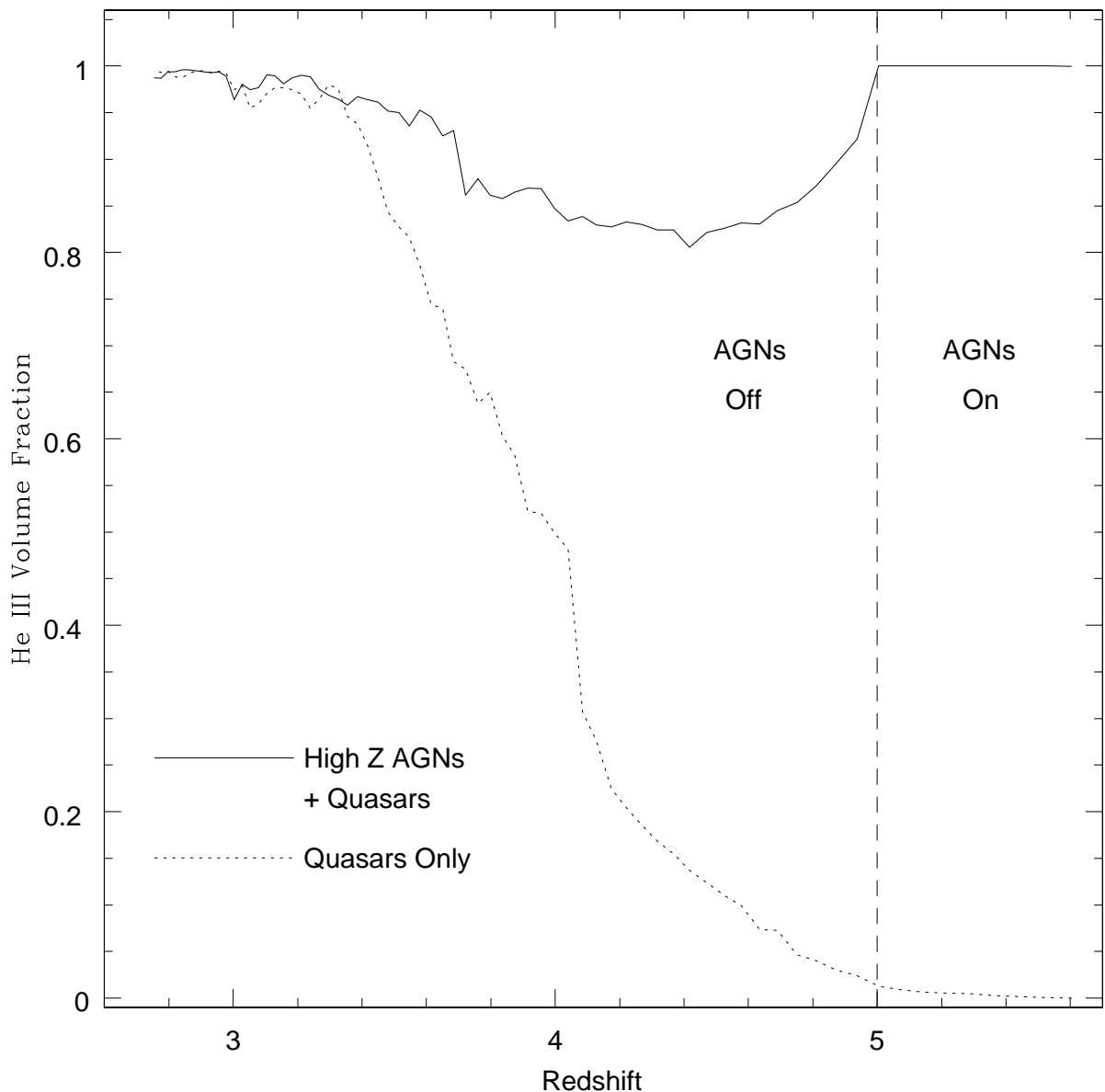}}
\end{picture}
\caption{Evolution of the He {\small III}
volume-weighted ionisation fraction due to quasars turning on at
$z<5$. The dotted line represents the results from the simulation
conducted by Sokasian et al. (2002) involving a realistic quasar model
(model 5). The solid line shows the resultant evolution in the
same simulation now involving the scenario where the IGM was first
highly ionised due to the presence of hard sources (such as AGNs)
which abruptly turn off at $z=5$. Note the ionisation fraction in the
latter case eventually converges to the same levels as in the case
involving quasars only, rendering it indistinguishable in light of
current observational results for He {\small II} opacities near
$z\simeq3.2$ (see Sokasian et al. 2002).}
\end{figure}

\clearpage
\begin{figure}[htb]
\figurenum{13}
\setlength{\unitlength}{1in}
\begin{picture}(6,6.5)
\put(1.00,-2.2){\includegraphics{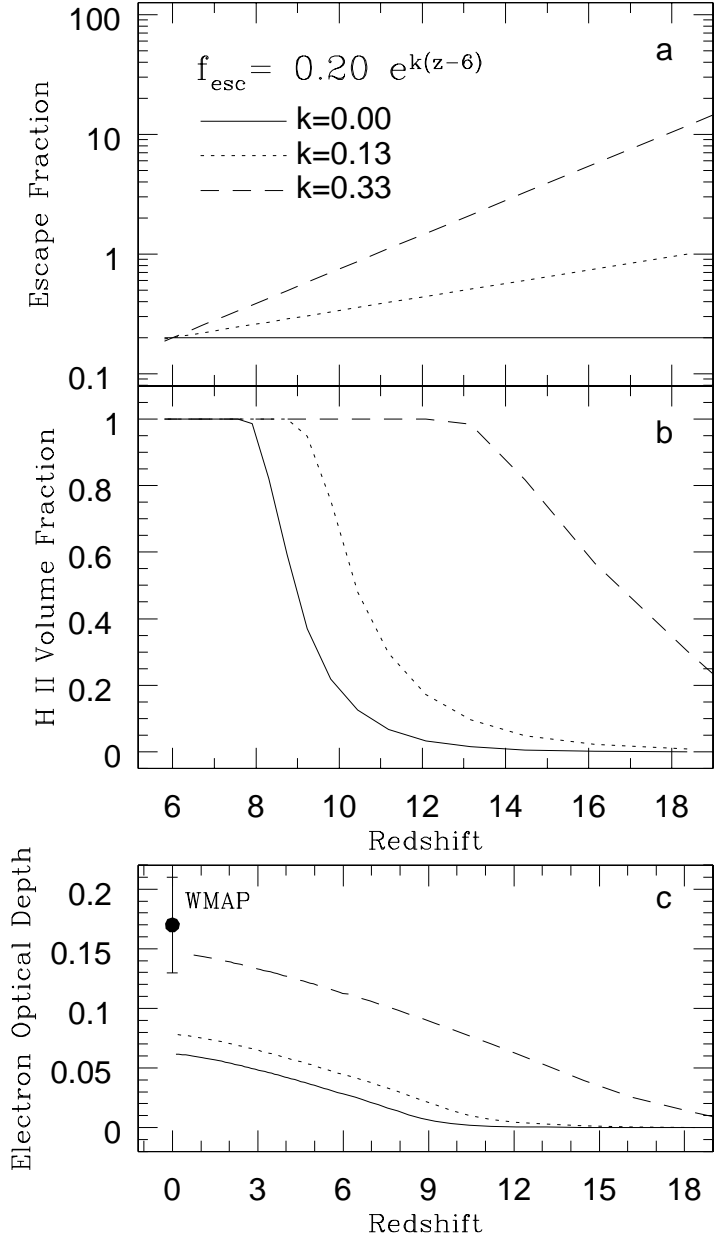}}
\end{picture}
\caption{Redshift evolution of the escape fraction (a), H {\small II}
volume-weighted ionisation fraction (b), and electron optical depth
(c) for the case with a constant escape fraction of $f_{esc}=0.20$
({\it solid-line}), and two cases where we have adopted an evolving
escape fraction of the form $f_{esc}(z)=0.20 e^{k(z-6)}$.  The case
with $k=0.13$ ({\it dotted-line}) represents the scenario where there is
an evolution of the escape fraction from 0.20 at $z=6$ (necessary to
provide a consistent match with the Becker et al. 2001 observations)
to unity at $z=18$ around when the first star-forming sources turn on. 
The case with $k=0.33$ results in an escape fraction which is larger than
unity beyond $z\gtrsim11$ and is meant to serve as an illustrative
example of the additional ionising flux necessary at high redshifts in
order to match the electron optical depth measurement implied by the
Kogut et al. (2003) ``model independent'' analysis (WMAP data point in
panel c).}
\end{figure}

\end{document}